\documentclass[10pt,journal]{IEEEtran}
\usepackage{float}
\usepackage{tabularx}
\usepackage{mathtools}
\usepackage{amsmath}
\usepackage{hhline}
\usepackage{silence}
\usepackage{graphicx}
\usepackage{soul}
\usepackage{subcaption}

\usepackage{xcolor}
\usepackage{multirow}
\usepackage{graphicx}
\usepackage{amsmath}
\usepackage[english]{babel}
\usepackage{booktabs}
\usepackage{array}
\usepackage{paralist}
\usepackage{threeparttable}
\usepackage{lipsum}
\usepackage{flushend}
\usepackage{cuted}
\usepackage{algpseudocode}
\usepackage{algorithm}
\usepackage{amssymb}
\usepackage{multicol}

\usepackage{cite}
\usepackage{makecell}
\usepackage{url}
\usepackage[table]{xcolor}
\usepackage[normalem]{ulem}
\usepackage{colortbl}
\usepackage[
    colorlinks=true,
    linkcolor=blue,   
    citecolor=blue,   
    urlcolor=blue     
]{hyperref}

\usepackage{enumitem}

\newcommand{\ignore}[1]{}

\newcommand{\blue}[1]{{\textcolor{blue}{#1}}}

\usepackage{listings}
\lstdefinestyle{promptstyle}{
  basicstyle=\ttfamily\footnotesize,   
  columns=fullflexible,
  keepspaces=true,
  breaklines=true,
  showstringspaces=false,
  frame=single,
  rulecolor=\color{promptgray},
  backgroundcolor=\color{yellow!10},
  xleftmargin=0.15em,
  xrightmargin=0.15em,
  aboveskip=0.15em,
  belowskip=0.15em,
  commentstyle=\color{promptgray},
  keywordstyle=\color{promptgray},
  stringstyle=\color{promptgray},
  escapeinside={(*@}{@*)} 
}

\definecolor{promptgray}{rgb}{0.5,0.5,0.5}
\definecolor{lightcream}{RGB}{255,250,235}

\definecolor{lightblue}{RGB}{242,247,252}

\definecolor{lightgray}{RGB}{248,248,248}

\definecolor{lightgreen}{RGB}{243,250,245}

\definecolor{lightpeach}{RGB}{255,246,240}

\definecolor{lightlavender}{RGB}{248,246,252}

\definecolor{lightcyan}{RGB}{242,250,250}

\definecolor{lightyellow}{RGB}{255,252,235}

\definecolor{gmidestimator_color}{RGB}{255,234,234}

\definecolor{equationgenerator_color}{RGB}{248,255,240}

\definecolor{equationrepair_color}{RGB}{240, 247, 255}

\definecolor{iter1bg}{RGB}{255,242,238}   
\definecolor{iter2bg}{RGB}{255,249,225}   
\definecolor{iter3bg}{RGB}{237,249,241}   

\definecolor{iter1frame}{RGB}{190,70,60}
\definecolor{iter2frame}{RGB}{190,145,25}
\definecolor{iter3frame}{RGB}{45,135,75}

\definecolor{failred}{RGB}{190,35,35}
\definecolor{passgreen}{RGB}{25,125,65}
\definecolor{darkblue}{RGB}{20,65,120}

\usepackage{adjustbox}

\usepackage{pifont}

\begin{document}
\pagestyle{empty}
\renewcommand{\footnote}[1]{}

\title{NEMESIS2.0: LLM-Aided Netlist-Driven Performance Equation Modeling for Accelerated Analog Circuit Sizing}
\author{
\IEEEauthorblockN{Subhadip Ghosh, Ramesh Harjani, and Sachin S. Sapatnekar}
\thanks{The authors are with the Department of Electrical and Computer Engineering at the University of Minnesota, Minneapolis, MN, USA. This work was supported in part by the NSF under award 2212345.}
}

\maketitle
\thispagestyle{empty}
\begin{abstract}
This work presents \textsc{NEMESIS2.0}, an open-source LLM-aided framework that automatically derives equation-based performance models for analog circuits. The resulting equations provide fast, accurate surrogates for SPICE evaluation. To validate the method, equation-based models are developed for seven circuit test cases in a commercial 65\,nm CMOS technology, covering 15 metrics on five operational transconductance amplifier topologies, and six metrics on two dynamic comparator topologies. When tested on 1000 unseen variants of these seven designs, each corresponding to a different set of independently sampled transistor sizes, the accuracy is found to be within 5\% of SPICE. To demonstrate use cases for NEMESIS2.0, the generated models are used as performance evaluators in multiple transistor sizers based on different population-based stochastic search optimizers. For each of the seven test cases, transistor sizing is performed for 100 different sets of performance specifications, with final SPICE simulation confirming that 83.3\% of the resulting designs meet all specifications after a single sizing run.  The rest fail to meet the specifications due to discrepancies between the model and SPICE, and require iterative resizing under tightened specifications; all such scenarios meet specifications within five iterations. Within the transistor sizers, this equation-based performance analysis evaluates $\approx$400 design candidates per second, versus 0.14 candidates per second for SPICE-based evaluation, achieving 196--238$\times$ sizing runtime speedup, including final SPICE verification.
\end{abstract}


\section{Introduction}
\label{sec:introduction}

\noindent
Analog integrated circuits interface physical signals with digital systems through sensing, amplification, and data conversion. Their performance depends on circuit topology, device dimensions, operating regions, parasitics, and loading conditions, which collectively influence multiple, often competing, performance metrics.
Meeting the desired specifications therefore requires systematic optimization of circuit parameters to navigate these performance tradeoffs. For a selected topology, schematic-level sizing determines transistor dimensions and bias conditions
to meet the design specifications. To automate this process, optimization algorithms iteratively propose candidate transistor sizes and use a performance evaluator to estimate performance metrics of the design. These estimates guide candidate selection and refinement until the objectives are met or the process times out.
The evaluator is therefore central to automated sizing: slow evaluation limits the number of candidates that can be explored within a fixed runtime budget, while inaccurate predictions can misguide the optimizer.

Several automated sizing approaches employ SPICE-based evaluation, including methods based on stochastic search~\cite{koza_96,Kruiskamp_95,liu_09_hybrid,gielen_90,rutenbar_96,vural_12,liu_09,nuno_12}, Bayesian optimization~\cite{lyu_18,lyu_18_2,zhang_20}, reinforcement learning~\cite{settaluri_20,zhao_20,choi_23_2}, and large language models (LLMs)~\cite{yin_24,lai_25}. SPICE provides accurate performance estimates, but it is computationally expensive, particularly over the large number of candidate sizes that are evaluated by sizing algorithms. To reduce this runtime cost, data-driven surrogates, trained on SPICE-generated samples, have been used to estimate circuit performance using response-surface models~\cite{rsm_16,rsm_08}, probabilistic models~\cite{lyu_18,lyu_18_2,zhang_20}, neural networks~\cite{budak_21,ghosh_2025}, and graph-based models~\cite{wang_2020}. 
Although they are faster than SPICE, these surrogates require extensive training data and adequate design-space coverage to achieve accurate predictions. Moreover, while SPICE and data-driven surrogates provide performance estimates, they do not inherently expose the analytical relationships between device parameters and circuit performance metrics. Such relationships can provide additional interpretability, allowing circuit designers to understand how individual device parameters influence performance, identify potential sources of performance degradation, and reason about design tradeoffs.

\begin{figure}[t]
    \centering
    \subfloat[\label{fig:highlevel_modeling}]{%
        \includegraphics[width=0.48\columnwidth]{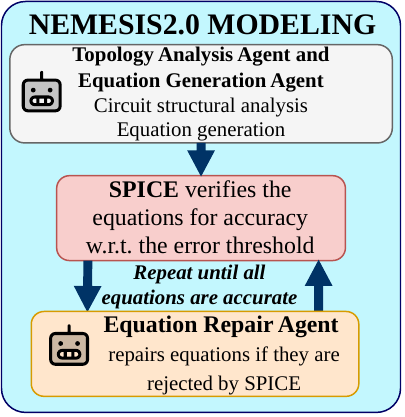}
    }
    \hfill
    \subfloat[\label{fig:highlevel_sizing}]{%
        \includegraphics[width=0.45\columnwidth]{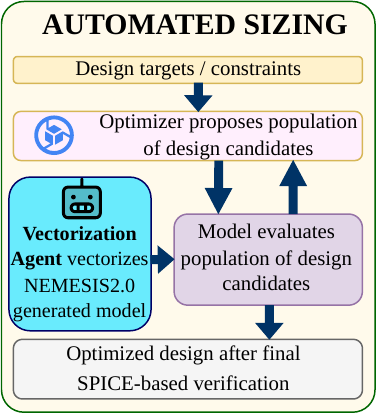}
    }
    \caption{(a) \textsc{NEMESIS2.0} performance modeling framework and its application in (b) the automated sizing framework.}
    \label{fig:highlevel_flow}
    \vspace{-8mm}
\end{figure}

Analytical modeling offers a way to explicitly capture these relationships without learning them primarily from sampled design--performance data.
Methods based on analytical equations~\cite{abel_22_2,graeb_01,graeb_08,harjani_89,koh_90,Gebru26}, symbolic analysis~\cite{gielen_90}, and geometric programming~\cite{daems_03,mandal_01,hershenson_98} follow this principle, providing explicit performance models that can be evaluated rapidly without requiring large training datasets. 
However, their reliance on simplified MOS models and operating-region-specific assumptions can compromise accuracy as device dimensions, bias conditions, and transistor operating regions change during sizing. Moreover, these equations are required for a vast range of circuits, each of which may have multiple topologies and differing performance metrics: deriving and adapting equations across circuit topologies and performance metrics requires substantial manual effort. These limitations motivate automatically generated equation-based performance models that retain the speed and structured, interpretable circuit-level representation of analytical methods while achieving SPICE-referenced accuracy across varying operating conditions, thereby providing fast evaluators for schematic-level sizing.

To address these challenges, we present \textsc{NEMESIS2.0}, an LLM-aided framework that extends our \textsc{NEMESIS} work~\cite{ghosh_26} to automatically generate SPICE-verified, interpretable performance equations across circuit topologies.
\ignore{ MOVING ALL COMPARISONS WITH PRIOR NEMESIS TO A DIFFERENT PARA ... AFTER THE CURRENT FRAMEWORK IS EXPLAINED IN DETAIL Unlike our earlier \textsc{NEMESIS} framework~\cite{ghosh_26}, which generated performance equations around a single operating-point configuration, this framework constructs and verifies each equation across multiple operating-point profiles spanning distinct speed, gain, and power tradeoffs. This broader operating-condition coverage improves model generality, enabling usage across different sizing objectives without regenerating the equations.} 
As shown in Figure~\ref{fig:highlevel_modeling}, the framework integrates circuit interpretation, operating-point selection, equation generation, and SPICE-guided refinement to construct reusable performance prediction models. We integrate these models as performance evaluators into an automated sizing framework (Figure~\ref{fig:highlevel_sizing}) where population-based optimization algorithms are used to optimize transistor dimensions and bias conditions under application-specific performance objectives and constraints. By using analytical equations for performance estimation, the framework reduces the computational cost of evaluating candidate designs without repeated metric-specific SPICE simulations, while retaining full SPICE verification of the final optimized designs.

The performance models in \textsc{NEMESIS2.0} are constructed through a sequential workflow primarily involving three LLM-based agents. First, the \textit{Topology Analysis Agent} interprets circuit connectivity to identify transistor roles and selects representative operating-point configurations. Next, the \textit{Equation Generation Agent} uses the identified circuit structure and operating-point configurations to derive analytical equations relating device-level electrical parameters to circuit performance metrics and evaluates their accuracy against SPICE simulations. Finally, the \textit{Equation Repair Agent} selectively refines equations that fail to meet the prescribed SPICE-referenced error threshold.

The \textit{Topology Analysis Agent} begins by analyzing the circuit netlist and schematic to identify transistor roles and functional primitives, such as differential pairs~(DPs) and current mirrors~(CMs). To capture variations in circuit performance across diverse transistor operating conditions, the agent selects multiple operating-point configurations using the transconductance-efficiency ratio ($g_m/I_D$), considering the functional role of each transistor. The $g_m/I_D$ ratio governs transistor inversion levels and influences tradeoffs among speed, intrinsic gain, power efficiency, and other performance characteristics~\cite{silviera_96,jespers_09,jespers_17}. Accordingly, the agent constructs three representative configurations targeting high-speed, balanced, and high-gain operation, with transistor operating regions selected according to their functional roles. These configurations provide diverse operating conditions for generating and verifying analytical equations, improving their robustness.

Using the circuit interpretation and operating-point configurations provided by the \textit{Topology Analysis Agent}, the \textit{Equation Generation Agent} constructs an analytical equation for each performance metric in terms of device-level small- and large-signal parameters, such as transconductance ($g_m$), output conductance ($g_{ds}$), drain current ($I_D$), and parasitic capacitances. Retrieval-augmented generation~(RAG)~\cite{lewis_20} supports equation construction by retrieving previously verified performance equations for similar circuits. To predict circuit performance using the generated equations, the required device parameters must be obtained. Accordingly, the \textit{Topology Analysis Agent} maps the assigned $g_m/I_D$ targets to realizable transistor dimensions using technology-characterized lookup tables~(LUTs). Rather than relying on simplified MOS device models~\cite{harjani_89,koh_90,Gebru26}, \textsc{NEMESIS2.0} extracts the device parameters through computationally inexpensive SPICE operating-point (\texttt{.OP}) analyses~\cite{ghosh_26}. The extracted parameters are then substituted into the generated equations to estimate circuit performance across all three operating-point configurations.

To assess the accuracy of the generated equations, their predicted performance metrics are compared against reference values obtained from metric-specific SPICE simulations across all three operating-point configurations. Equations that meet the prescribed error threshold are retained, while those that exceed the threshold are forwarded to the \textit{Equation Repair Agent} for refinement. Guided by the verification feedback, the agent adds or removes relevant small- and large-signal parameters based on LLM reasoning to improve equation accuracy while preserving the underlying circuit relationships. The verification and refinement process continues until the equations satisfy the error threshold across all three configurations, yielding one SPICE-verified analytical equation per performance metric.

\begin{figure*}[t]
    \centering
    \includegraphics[width=\textwidth]{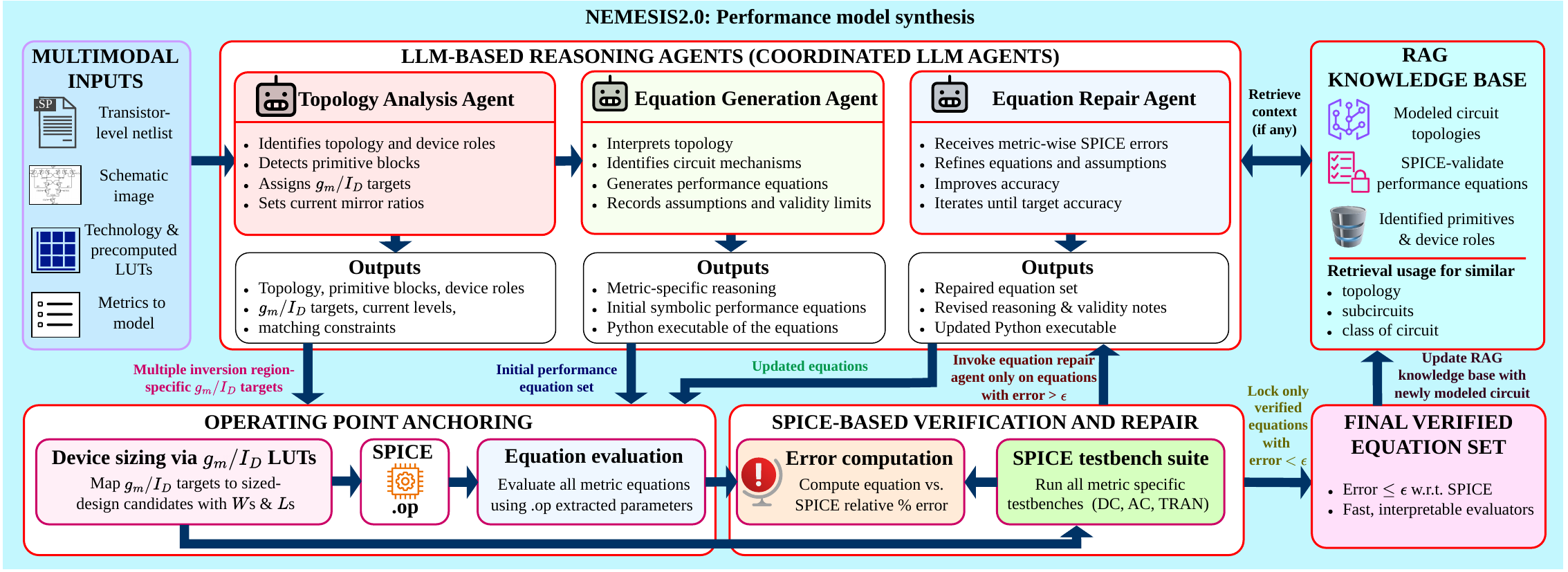}
    \vspace{-4mm}
    \caption{Overview of the \textsc{NEMESIS2.0} performance modeling framework. }
    \label{fig:nemesis_2.0}
    \vspace{-5mm}
\end{figure*}

The proposed modeling workflow introduces two key extensions over our earlier \textsc{NEMESIS} framework~\cite{ghosh_26}. First, whereas \textsc{NEMESIS} generated performance equations around a single application-specific operating-point configuration, \textsc{NEMESIS2.0} constructs and verifies equations across three representative configurations spanning different performance tradeoffs. This broader coverage improves equation robustness across diverse transistor operating conditions. Second, while \textsc{NEMESIS} employed a single agent for equation generation and repair, \textsc{NEMESIS2.0} separates these responsibilities into dedicated agents, enabling targeted refinement based on SPICE verification feedback. Together, these extensions improve the generality of the performance equations and provide a more structured approach to their generation and refinement.

To demonstrate the applicability of the generated equations to automated circuit sizing, we construct performance models for five operational transconductance amplifier~(OTA) and two dynamic comparator~(COMP) topologies and integrate them into a population-based sizing framework. As a postprocessing task, an LLM-based \textit{Vectorization Agent} converts the verified analytical equations into executable vectorized models, enabling efficient performance prediction for multiple candidate sizing configurations. To assess their portability across optimization methods, we independently employ three algorithms: genetic algorithm~(GA)~\cite{holland_75}, particle swarm optimization~(PSO)~\cite{kennedy_95}, and differential evolution~(DE)~\cite{storn_97}. Each algorithm optimizes transistor dimensions and bias conditions for a fixed circuit topology under the same performance objectives and constraints, using the equation-based evaluators. Finally, the optimized designs are verified against the original target specifications through SPICE simulations.

The key contributions of this work are:

\begin{itemize}[leftmargin=*, labelsep=0.5em]

\item \textbf{Robust performance models:} Analytical equations relate transistor electrical parameters to up to 15 circuit performance metrics, meeting a $5\%$ SPICE-referenced error threshold capturing speed, gain, and power tradeoffs.

\item \textbf{Automated modeling across circuit topologies:} Dedicated LLM-based agents automate performance modeling and targeted refinement, demonstrated on five OTA and two COMP topologies without manual equation derivation.

\item \textbf{High-throughput performance evaluation:} SPICE-verified equations enable accurate performance estimation for $\approx$400 candidate designs per second, avoiding repeated metric-specific SPICE simulations during optimization.

\item \textbf{Specification-compliant automated sizing:} Across seven circuits, each sized for 100 different specification sets, $83.3\%$ of designs meet all targets on the first attempt in $<$35sec; the rest succeed within five iterative sizing attempts.

\item \textbf{Accelerated automated sizing:} By avoiding repeated metric-specific SPICE simulations during optimization, the framework achieves 196$\times$--238$\times$ speedups over SPICE-in-the-loop sizing for single-attempt cases and 39$\times$--50$\times$ for five-attempt cases, while retaining final SPICE verification.

\end{itemize}
The remainder of this paper is organized as follows. Section~\ref{sec:NEMESIS2.0_modeling} presents the NEMESIS2.0 performance-modeling framework, Section~\ref{sec:agentic_sizing} describes its application in the agentic sizing framework, and Section~\ref{sec:experiments_results} presents the experimental results. Section~\ref{sec:conclusion} concludes the paper.

\vspace{-3mm}

\section{\textsc{NEMESIS2.0} Modeling Framework}
\label{sec:NEMESIS2.0_modeling}

\noindent
Figure~\ref{fig:nemesis_2.0} illustrates the complete modeling workflow, from circuit interpretation to the generation of SPICE-verified performance equations. The flow begins with the \textit{Topology Analysis Agent}, which processes the circuit netlist and schematic to identify primitives, transistor roles, and signal paths, and establishes three operating-point profiles with assigned $g_m/I_D$ values, branch currents, and channel lengths. Guided by this circuit interpretation, the operating-point profiles, and relevant knowledge retrieved from the RAG knowledge base, the \textit{Equation Generation Agent} constructs analytical equations and their executable implementations for each performance metric.

\begin{figure}[t]
    \centering

    \resizebox{0.95\columnwidth}{!}{%
    \begin{minipage}{\columnwidth}
        \centering

        \begin{subfigure}[c]{0.58\columnwidth}
            \centering
            \includegraphics[width=\linewidth]{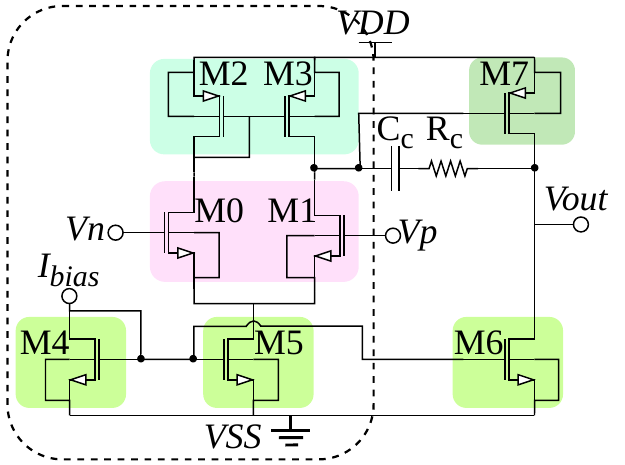}
            \caption{}
            \label{fig:2sota}
        \end{subfigure}
        \hfill
        \begin{subfigure}[c]{0.39\columnwidth}
            \centering
            \includegraphics[width=\linewidth]{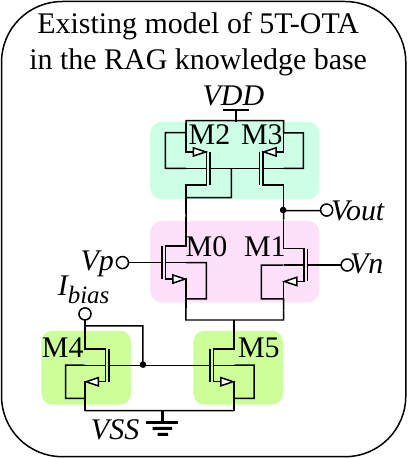}
            \caption{}
            \label{fig:5tota}
        \end{subfigure}

    \end{minipage}%
    }
    \caption{(a) Schematic of the 2S-OTA; the dashed boundary encloses the first stage, corresponding to (b) the previously modeled 5T-OTA from the RAG knowledge base. 
    }
    \label{fig:2sota_5tota}
    \vspace{-5mm}
\end{figure}

For each profile, the operating-point anchoring stage uses technology-characterized LUTs to map the assigned $g_m/I_D$ values and channel lengths to transistor current densities, from which transistor widths are determined using the branch currents. SPICE \texttt{.OP} analyses then provide the device parameters needed to evaluate the metric equations, while metric-specific DC, AC, and transient simulations establish reference performance values. The verification stage compares equation predictions against these SPICE references across all three profiles, retaining equations that satisfy the prescribed error threshold and forwarding inaccurate equations to the \textit{Equation Repair Agent}. The agent iteratively refines the equations while reusing previously extracted device parameters and SPICE reference values, avoiding redundant simulations. Once verified, the equations and their executable implementations are stored in the RAG knowledge base for future reference. The following subsections describe each stage of the flow in detail, using a two-stage OTA (2S-OTA) (Figure~\ref{fig:2sota}) comprising a five-transistor OTA (5T-OTA) (Figure~\ref{fig:5tota}) first stage and a common-source (CS) second stage as a running example.
\vspace{-2mm}

\subsection{Multimodal circuit interpretation and knowledge retrieval}
\label{sec:multimodal_rag}

\noindent
The flow begins with the multimodal-input block in Figure~\ref{fig:nemesis_2.0}, where the transistor-level netlist and schematic are jointly analyzed to identify circuit structures and device roles. The netlist specifies device connectivity, terminal assignments, model types, and parameters, while the schematic provides visual cues for symmetry, multiple stages, and signal flow that may be less apparent from connectivity alone. Together, these inputs enable the \textit{Topology Analysis Agent} to identify circuit topology, functional primitives, and device roles, and construct operating-point profiles. The resulting circuit interpretation also guides the \textit{Equation Generation Agent} in identifying the physical mechanisms relevant to each performance metric. The significance of incorporating schematic information alongside the netlist is further examined in Section~\ref{sec:multimodal_input}.

To complement circuit interpretation with knowledge from previously modeled circuits, we represent the transistor-level netlist as a terminal-labeled bipartite graph,
$\mathcal{G}=(\mathcal{V}_{\mathrm{dev}}\cup\mathcal{V}_{\mathrm{net}},\mathcal{E},\ell_v,\ell_e)$,
where $\mathcal{V}_{\mathrm{dev}}$ and $\mathcal{V}_{\mathrm{net}}$ denote device and net vertices, and $\mathcal{E}$ denotes the edges connecting device terminals to nets respectively. The vertex labels $\ell_v$ distinguish device types and identify nets connected to external circuit pins, while the edge labels $\ell_e$ preserve device-terminal roles such as drain, gate, source, and bulk. External circuit pins retain functional labels identifying signal inputs, outputs, power supplies, and bias terminals. In contrast, transistor instance names and internal net names are excluded because they do not affect circuit connectivity or device-terminal relationships. Device dimensions are also omitted to represent circuit structure independently of a particular sizing configuration. The resulting graph therefore preserves terminal-aware connectivity and external pin functionality while remaining insensitive to changes in device naming, internal net naming, and transistor sizing. The RAG knowledge base stores these representations alongside previously modeled circuit information and SPICE-verified performance equations.

To retrieve structurally related circuits, we evaluate similarity using two complementary descriptors: a terminal-aware Weisfeiler--Lehman (WL) fingerprint~\cite{wl_11} capturing local device-to-net connectivity and an analog-motif vector capturing functional structures. The latter records device composition, diode-connected devices, shared-gate groups, differential-pair and current-mirror patterns, and output, supply, and bias connections. For a query circuit $q$ and stored circuit $i$, the similarity score is
$
S(q,i)=
\alpha s_{\mathrm{WL}}(q,i)
+(1-\alpha)s_{\mathrm{motif}}(q,i),
\label{eq:rag_similarity}
$
where $s_{\mathrm{WL}}$ is the multiset-Jaccard similarity~\cite{jaccard_1901} between WL fingerprints and $s_{\mathrm{motif}}$ is the cosine similarity between motif vectors. We heuristically set $\alpha=0.6$ for all queries, assigning slightly greater weight to terminal-level connectivity while retaining information about higher-level circuit structures. Stored circuits are ranked according to $S(q,i)$, and the top $N_{\mathrm{ret}}=3$ records are supplied to the \textit{Equation Generation Agent}.

Figure~\ref{fig:2sota_5tota} illustrates this retrieval process using the 2S-OTA and the previously modeled 5T-OTA. The colored blocks highlight the circuit primitives identified by our approach and are not part of the input. The shared 5T-OTA first stage provides functional similarities captured by the analog-motif descriptor, whereas the additional CS stage and compensation network alter the connectivity and reduce the WL similarity. The combined similarity score therefore retrieves the 5T-OTA as a related, but nonidentical, circuit. Its stored structural information and SPICE-verified equations provide references for modeling the shared first stage. Using these references together with the target circuit interpretation, the \textit{Equation Generation Agent} accounts for the additional CS stage, compensation elements (R$_\text{c}$ and C$_\text{c}$), and modified signal and loading paths in the 2S-OTA. Thus, retrieved equations guide the analytical derivation without being transferred directly to the target circuit. We independently verify the generated equations against SPICE for the 2S-OTA, and only verified equations are added to the RAG knowledge base.

\begin{figure}[t]
    \centering
    \begin{minipage}{\columnwidth}
    \centering

    \begin{minipage}[t]{0.4\linewidth}
\begin{lstlisting}[style=promptstyle,basicstyle=\ttfamily\fontsize{5.8}{5.7}\selectfont,breaklines=true,columns=fixed,
basewidth=0.52em, stringstyle=\color{promptgray}, escapeinside={(*@}{@*)},backgroundcolor=\color{gmidestimator_color}]
(*@\textbf{"num\_operating\_points"}@*):3,
(*@\textbf{"topology\_analysis"}@*):{
 "class":"OTA",
 "topology":"two-stage miller compensated OTA"},
 
(*@\textbf{"primitive\_groups"}@*):{
 "input_pair":[M0,M1],
 "tail":[M5],
 "current_mirror":[[M4,M5,M6],  [M2,M3]],
 "active_load":[M2,M3],
 "output_stage":[M7,M6],
 "bias_reference":[M4],
 "matched_devices":[[M0,M1],    [M2,M3]]},

(*@\textbf{"gmid\_groups"}@*):{
 (*@\textbf{"G1"}@*):[M0,M1],
 (*@\textbf{"G2"}@*):[M2,M3],
 (*@\textbf{"G3"}@*):[M4,M5,M6],
 (*@\textbf{"G4"}@*):[M7]},
 
(*@\textbf{"device\_roles"}@*):{
 "M0":"negative-input device",
 "M1":"positive-input device",
 "M2":"PMOS mirror reference",
 "M3":"PMOS active load",
 "M4":"NMOS bias reference",
 "M5":"tail current source",
 "M6":"second-stage sink",
 "M7":"second-stage PMOS"}
\end{lstlisting}
    \end{minipage}
    \hfill
    \begin{minipage}[t]{0.57\linewidth}
\begin{lstlisting}[style=promptstyle,basicstyle=\ttfamily\fontsize{5.8}{5.7}\selectfont,breaklines=true,columns=fixed,
basewidth=0.52em, stringstyle=\color{promptgray}, escapeinside={(*@}{@*)},backgroundcolor=\color{gmidestimator_color}]
(*@\textbf{Operating-Point Profile 1}@*):{
 (*@\textbf{"intent":"high\_speed"}@*),
 (*@\textbf{"G1"}@*):{"gmid":9,"region":"strong",
 "basis":"higher input bandwidth"},
 (*@\textbf{"G2"}@*):{"gmid":9,"region":"strong",
 "basis":"reduced 1st-stage pole capacitance"},
 (*@\textbf{"G3"}@*):{"gmid":11,"region":"moderate",
 "basis":"improved output resistance"},
 (*@\textbf{"G4"}@*):{"gmid":6,"region":"strong",
 "basis":"strong Miller-output drive"}},

(*@\textbf{Operating-Point Profile 2}@*):{
 (*@\textbf{"intent":"balanced",}@*)
 (*@\textbf{"G1"}@*):{"gmid":16,"region":"moderate",
 "basis":"efficiency-speed balance"},
 (*@\textbf{"G2"}@*):{"gmid":12,"region":"moderate",
 "basis":"mirror gain and capacitance"},
 (*@\textbf{"G3"}@*):{"gmid":12,"region":"moderate",
 "basis":"output resistance and compliance"},
 (*@\textbf{"G4"}@*):{"gmid":9,"region":"strong", 
 "basis":"output drive"}},

(*@\textbf{Operating-Point Profile 3}@*):{
 (*@\textbf{"intent":"high\_gain",}@*)
 (*@\textbf{"G1"}@*):{"gmid":18,"region":"weak",
 "basis":"high input gm efficiency"},
 (*@\textbf{"G2"}@*):{"gmid":16,"region":"moderate",
 "basis":"Improved load resistance"},
 (*@\textbf{"G3"}@*):{"gmid":15,"region":"moderate",
 "basis":"current-source output resistance"},
 (*@\textbf{"G4"}@*):{"gmid":10,"region":"moderate",
 "basis":"pole separation and drive"}}
\end{lstlisting}
    \end{minipage}

    \end{minipage}

\caption{\textit{Topology analysis agent} output of 2S-OTA showing topology interpretation and assigned operating-point profiles.}
    \label{fig:2sota_gmid_output}
    \vspace{-5mm}
\end{figure}

\vspace{-3mm}
\subsection{Multi-region $g_m/I_D$ operating-point profile generation}
\label{sec:multiregion_modeling}

\noindent
Building on the circuit interpretation and retrieved knowledge, the \textit{Topology Analysis Agent} constructs three representative operating-point profiles to capture variations in circuit behavior across transistor inversion regions. Changes in inversion region affect device-level electrical characteristics and their contributions to circuit performance, making it important to account for these variations when developing analytical models. The agent generates high-speed, balanced, and high-gain profiles by assigning different combinations of strong, moderate, and weak inversion to functional transistor groups. These profiles provide distinct operating conditions for evaluating the analytical equations without exhaustively exploring all possible operating-region combinations.

To construct these profiles, the agent first groups transistors according to their functional roles and shared operating-point constraints rather than circuit stages alone. The grouping is derived from the supplied netlist, schematic, and prompt instructions, with additional guidance from structurally related circuits retrieved through the RAG mechanism described in Section~\ref{sec:multimodal_rag}. Retrieved records, when available, provide reference primitive classifications, connectivity relationships, and $g_m/I_D$ assignments; otherwise, the agent determines these from the target circuit information. The left column of Figure~\ref{fig:2sota_gmid_output} illustrates the resulting grouping for the 2S-OTA: G1 comprises the input differential pair, G2 the PMOS active-load current mirror, G3 the NMOS bias-derived current-mirror devices, and G4 the second-stage gain device. Although M6 belongs to the second stage, it is grouped with M4 and M5 in G3 because these devices share a bias-derived $g_m/I_D$ constraint. Within these groups, matched devices receive equal $g_m/I_D$ targets, while current-mirror devices additionally share channel lengths, with their branch currents determined by the corresponding mirror ratios.

To assign numerical $g_m/I_D$ targets, we provide the agent with inversion-region ranges and primitive-specific sizing guidelines derived from established $g_m/I_D$ methodologies~\cite{silviera_96,jespers_17}. The agent selects values appropriate to the target technology, using assignments from retrieved circuits as additional references when available. The right column of Figure~\ref{fig:2sota_gmid_output} shows the resulting profiles for the 2S-OTA. In the high-speed profile, G1 and G2 use $g_m/I_D=9$, and G4 uses $g_m/I_D=6$, placing these groups in strong inversion, while G3 operates in moderate inversion at $g_m/I_D=11$. The balanced profile places G1--G3 in moderate inversion while retaining G4 in strong inversion for output drive. In the high-gain profile, G1 shifts to weak inversion at $g_m/I_D=18$, while G2--G4 operate at higher $g_m/I_D$ values within moderate inversion. These assignments vary device operating conditions while preserving the circuit's functional and biasing relationships. The resulting $g_m/I_D$ targets, branch currents, and channel lengths are subsequently mapped to transistor dimensions using technology-characterized lookup tables~(LUTs). SPICE operating-point analyses of the resulting profiles then provide distinct sets of device parameters for evaluating the same analytical equations under different operating conditions.
\vspace{-5mm}

\subsection{Performance equation generation}
\label{sec:equation_generation}

\noindent
The \textit{Equation Generation Agent} constructs topology-specific analytical performance equations using the circuit interpretation, operating-point profiles, and the retrieved knowledge from the RAG knowledge base. For each requested performance metric $m$, the agent generates a single equation that is evaluated across all three operating-point profiles:
\begin{equation}
\widehat{y}_{m,k} =
f_{m,\mathcal{N}}
\!\left(
\{\mathbf{x}_{d,k}\}_{d\in\mathcal{D}},\mathbf{u}
\right),
\label{eq:eqgen_mapping}
\end{equation}
where $f_{m,\mathcal{N}}$ denotes the equation for metric $m$ and netlist $\mathcal{N}$, and $\widehat{y}_{m,k}$ is its prediction at operating-point profile $k$. The set $\mathcal{D}$ contains the circuit transistors, while $\mathbf{x}_{d,k}$ represents the profile-dependent information for transistor $d$. This includes device dimensions---width ($W$) and channel length ($L$); DC operating-point quantities---terminal currents and voltages, threshold voltage ($V_{\mathrm{th}}$), and saturation voltage ($V_{\mathrm{DSAT}}$); and small-signal parameters---transconductance ($g_m$), output conductance ($g_{ds}$), body transconductance ($g_{mb}$), and intrinsic capacitances. The vector $\mathbf{u}$ contains external circuit and testbench parameters, such as supply voltages and load capacitance, that remain fixed across the three profiles.

To construct these equations, we prompt the agent to identify the devices, signal paths, and circuit relationships governing each performance metric in the target topology. Retrieved equations provide references for relevant analytical relationships, but the agent is instructed to account for differences in transistor connectivity, device functions, and loading conditions before adapting them to the target circuit. To reduce the risk of hallucinations, we further instruct the agent to avoid introducing devices, nodes, or circuit parameters that are absent from the input netlist.

\noindent
\textbf{Example: 2S-OTA.}
Representative equations for the 2S-OTA illustrate how the \textit{Equation Generation Agent} translates circuit interpretation into analytical performance models. For illustration, we compare the generated DC-gain equation with a conventional first-order model. Assuming matched devices and an ideal current mirror, the manual model approximates the overall gain as the product of the two stage gains~\cite{allen_holberg_2011} as:
\begingroup
\fontsize{8}{12}\selectfont
$$
\text{DC}\_\text{Gain}^{\mathrm{manual}}
\approx
20\log_{10}\left|
\frac{g_{m,M1}}{g_{ds,M1}+g_{ds,M3}}
\cdot
\frac{g_{m,M7}}{g_{ds,M6}+g_{ds,M7}}
\right|.
$$
\endgroup
Using the identified differential pair (M0--M1), PMOS current mirror (M2--M3), and second-stage common-source amplifier (M7), the agent instead generates:
\begin{lstlisting}[
style=promptstyle,
basicstyle=\ttfamily\fontsize{5.6}{5}\selectfont,
breaklines=true,
columns=fullflexible,
breakatwhitespace=false,
escapeinside={(*@}{@*)},
backgroundcolor=\color{equationgenerator_color},
aboveskip=2pt,
belowskip=0pt]
(*@\textbf{DC\_Gain}@*): 20*log10(abs(0.5*(gm_M1+gm_M0*gm_M3/(gm_M2+gds_M2))/(gds_M1+gds_M3)*(gm_M7/(gds_M6+gds_M7))))
(*@\textbf{Reasoning}@*): Differential current from M1 and the M0 current mirrored through M2-M3 drive the first-stage output resistance, followed by the gain of the M7 stage.
\end{lstlisting}
The generated equation separately captures the direct contribution of M1 and the contribution of M0 transferred through the CM, as indicated by the agent reasoning. The factor $g_{m,M3}/(g_{m,M2}+g_{ds,M2})$ accounts for unequal mirror $g_m$ and the finite conductance of the diode-connected M2. Under matched-device and ideal-mirror assumptions, the expression reduces to the conventional model. This comparison illustrates how the agent incorporates physically meaningful device-level effects without manual derivation.

The 3-dB-bandwidth (BW\_3dB) equation is derived from the dominant pole at the first-stage output, accounting for the drain capacitances of M1 and M3, the gate capacitance of M7, and the Miller-multiplied compensation capacitance:
\begin{lstlisting}[
style=promptstyle,
basicstyle=\ttfamily\fontsize{5.6}{5}\selectfont,
breaklines=true,
columns=fullflexible,
breakatwhitespace=false,
escapeinside={(*@}{@*)},
backgroundcolor=\color{equationgenerator_color},
aboveskip=2pt,
belowskip=0pt]
(*@\textbf{BW\_3dB}@*): (gds_M1+gds_M3)/(2*pi*(cdd_M1+cdd_M3+cgg_M7+Cc*(1+abs(gm_M7/(gds_M6+gds_M7)))))
(*@\textbf{Reasoning}@*): 1st-stage dominant pole includes M7 gate capacitance, M1/M3 drain capacitances, and Miller-multiplied Cc through the 2nd-stage gain.
\end{lstlisting}
For the output common-mode range (OCMR), the agent uses the saturation constraints of the output-stage NMOS device M6 and PMOS device M7 to determine the lower and upper output-voltage limits:

\begin{lstlisting}[
style=promptstyle,
basicstyle=\ttfamily\fontsize{5.6}{5}\selectfont,
breaklines=true,
columns=fullflexible,
breakatwhitespace=false,
escapeinside={(*@}{@*)},
backgroundcolor=\color{equationgenerator_color},
aboveskip=2pt,
belowskip=0pt]
(*@\textbf{OCMR\_Min/Max}@*): [Vss+abs(vdsat_M6), Vdd-abs(vdsat_M7)]
(*@\textbf{Reasoning}@*): M6 and M7 saturation boundaries determine the minimum and maximum output voltages, respectively.
\end{lstlisting}

This 2S-OTA example illustrates how the agent constructs initial analytical equations for different performance metrics from the identified circuit structure, with the remaining metrics modeled similarly. Each equation is accompanied by a Python-executable implementation that accepts device and testbench parameters to estimate performance. At this stage, the equations remain unverified and are subsequently evaluated against SPICE and selectively repaired, as described in Section~\ref{sec:verification_repair}.
\vspace{-3mm}


\definecolor{repairbg}{RGB}{255,250,232}
\definecolor{verifybg}{RGB}{244,248,252}
\definecolor{promptblue}{RGB}{20,65,120}
\definecolor{failred}{RGB}{190,35,35}
\definecolor{passgreen}{RGB}{25,125,65}
\definecolor{skipgray}{RGB}{105,105,105}

\begin{figure}[t]
\centering

\begin{minipage}[t]{0.38\columnwidth}
\centering
{\scriptsize\bfseries Representative repair prompt}
\vspace{1pt}

\begin{lstlisting}[
basicstyle=\ttfamily\fontsize{6}{5.6}\selectfont,
frame=single,
rulecolor=\color{promptblue},
backgroundcolor=\color{lightgreen},
breaklines=true,
breakatwhitespace=true,
breakautoindent=false,
breakindent=0pt,
columns=fixed,
basewidth=0.45em,
keepspaces=true,
showstringspaces=false,
escapeinside={(*@}{@*)},
xleftmargin=1pt,
xrightmargin=1pt,
aboveskip=0pt,
belowskip=0pt
]
(*@\textbf{GOAL:}@*)
Refine failed equations to within 5% of SPICE across all operating profiles.

(*@\textbf{DIAGNOSE AND REPAIR:}@*)
- Compare model and SPICE results across all OPs; use signed error and its variation to identify wrong estimation and region- dependent failures.
- Re-examine connectivity, physical signal paths, and the terms used in failed equations.
- Identify missing, incorrectly combined, or unjustified physics, effects arising from changes in bias, inversion, or headroom.
- Trace every correction to the netlist and SPICE .OP- extracted device parameters.
- Revise only FAILED equations, preserve PASSED equations, and re-evaluate the revised equations.

(*@\textbf{CONSTRAINTS:}@*)
- Use only netlist devices and permitted .OP device parameters.
- Every equation term must be physically traceable.
- No invented parameters, hidden nodes, empirical factors, or curve fitting.
- Do not repeat previously failed equations.
- Add only the minimum physical complexity needed to resolve the discrepancy.
\end{lstlisting}
\end{minipage}
\hfill
\begin{minipage}[t]{0.58\columnwidth}
\centering
\definecolor{color1}{RGB}{145,248,252}
{\scriptsize\bfseries Model vs. SPICE detailed summary}
\vspace{1pt}
\begin{lstlisting}[
basicstyle=\ttfamily\fontsize{5.8}{6}\selectfont,
frame=single,
rulecolor=\color{promptblue},
backgroundcolor=\color{color1},
columns=fixed,
basewidth=0.5em,
keepspaces=true,
escapeinside={(*@}{@*)},
showstringspaces=false,
xleftmargin=0pt,
xrightmargin=0pt,
aboveskip=0pt,
belowskip=0pt,
literate=
 {FAIL}{{\textcolor{failred}{FAIL}}}4
]
+------------+----------+----------+----------+
| UGF        | OP1      | OP2      | OP3      |
+------------+----------+----------+----------+
| SPICE (Hz) | 1.71e7   | 3.36e7   | 5.71e7   |
| Model (Hz) | 2.04e7   | 4.41e7   | 7.96e7   |
| Error      | 19.4%    | 31.1%    | 39.4%    |
| Status     | FAIL     | FAIL     | FAIL     |
+------------+----------+----------+----------+
(*@\textbf{History of FAILED equations}@*) over prior
repair iterations for the FAILED metrics only
\end{lstlisting}
\vspace{12pt}
\begin{lstlisting}[
basicstyle=\ttfamily\fontsize{5.8}{5.3}\selectfont,
frame=single,
rulecolor=\color{promptblue},
backgroundcolor=\color{lightyellow},
columns=fixed,
basewidth=0.5em, 
showstringspaces=false,
xleftmargin=-1pt,
xrightmargin=-2pt,
escapeinside={(*@}{@*)},
aboveskip=0pt,
belowskip=0pt,
literate=
 {PASS}{{\textcolor{passgreen}{PASS}}}4
 {FAIL}{{\textcolor{failred}{FAIL}}}4
 {SKIP}{{\textcolor{skipgray}{SKIP}}}4
]
(*@\textbf{MODEL VS. SPICE ERROR SUMMARY}@*)
+-------------+--------+--------+--------+------+
| Metric      |  OP1   |  OP2   |  OP3   |Status|
+-------------+--------+--------+--------+------+
|ICMR_Min     | -30.7% |  -7.3% | -17.5% | FAIL |
|ICMR_Max     | +47.1% | +50.0% | +52.6% | FAIL |
|DC_Gain      |  +0.2% |  +0.1% |  +0.1% | PASS |
|UGF          | +19.4% | +31.1% | +39.4% | FAIL |
|Phase_Margin | -92.1% | -81.0% | -67.0% | FAIL |
|BW_3dB       | +17.4% | +17.0% | +14.9% | FAIL |
|OCMR_Min     | +40.1% | +46.3% | +33.9% | FAIL |
|OCMR_Max     | +15.9% | +16.8% | +18.6% | FAIL |
|Slew_Rate+   | -64.4% | -40.0% | -61.8% | FAIL |
|Slew_Rate-   | -31.2% | -33.6% | -56.9% | FAIL |
|PSRR+        | -34.0% | -33.2% | -40.3% | FAIL |
|PSRR-        | -18.4% |  -9.8% |  -2.9% | FAIL |
|CMRR         | -25.0% | -11.1% | -21.8% | FAIL |
|Power        |  -0.0% |  -0.0% |  -0.0% | PASS |
+-------------+--------+--------+--------+------+
(*@\textbf{REPAIR RULE:}@*)
Use the complete error trend across all the
operating profiles. PASS models remain locked; 
each repaired equation must satisfy the error 
threshold for all OPs.
\end{lstlisting}
\end{minipage}

\caption{Representative inputs to the \textit{Equation Repair Agent}: (left) metric- and topology-agnostic repair instructions; (right-top) representative actual SPICE and model values supplied for a failed metric; and (right-bottom) signed model vs. SPICE error summary across the operating profiles for all the metrics.}
\label{fig:repair_prompt_context}
\vspace{-8mm}
\end{figure}

\subsection{SPICE-anchored equation verification and repair}
\label{sec:verification_repair}

\subsubsection{\textbf{SPICE-based equation verification}}
\label{sec:spice_verification}

\noindent
Although the generated equations capture physically motivated circuit relationships, they may omit relevant device contributions, loading effects, or parasitic capacitances, leading to inaccuracies across operating conditions. As in the original \textsc{NEMESIS} framework~\cite{ghosh_26}, we assess their accuracy by comparing equation predictions against reference values obtained from metric-specific DC, AC, and transient SPICE simulations. In \textsc{NEMESIS2.0}, we extend this verification across the three $g_m/I_D$ operating-point profiles---high-speed (OP1), balanced (OP2), and high-gain (OP3)---by evaluating the same analytical equation using the device parameters corresponding to each profile. The metric-specific testbenches are described in Section~\ref{sec:experiments_results}.

To reduce inaccuracies associated with simplified transistor models, we evaluate the analytical equations using device parameters extracted using \texttt{.OP} analysis from SPICE at the corresponding operating points. Consider, for example, a resistively loaded NMOS common-source stage, for which a candidate low-frequency voltage-gain ($A_v$) relationship is
\begingroup
\fontsize{8}{12}\selectfont
\begin{equation}
A_v\approx-\frac{g_m}{g_{\mathrm{out}}},
\qquad
g_{\mathrm{out}}=g_{ds}+\frac{1}{R_L},
\label{eq:cs_gain}
\end{equation}
\endgroup
where $R_L$ is the load resistance. Conventional first-order MOS analysis may approximate the device parameters as
\begingroup
\fontsize{8}{12}\selectfont
\begin{equation}
g_m\approx\frac{2I_D}{V_{GS}-V_{\mathrm{th}}},
\qquad
g_{ds}\approx\lambda I_D,
\label{eq:long_channel_params}
\end{equation}
\endgroup
where $V_{GS}-V_{\mathrm{th}}$ is the gate overdrive voltage, and $\lambda$ is the channel-length-modulation parameter. These approximations rely on long-channel, strong-inversion operation in saturation and typically treat $\lambda$ as constant. Their accuracy can therefore deteriorate as transistor dimensions and bias conditions change, particularly across different inversion regions.

The gain relationship in Eq.~\eqref{eq:cs_gain} illustrates the form of an analytical equation generated by the LLM. Rather than approximating $g_m$ and $g_{ds}$ using Eq.~\eqref{eq:long_channel_params}, we evaluate the same circuit-level gain relationship using the values extracted from SPICE \texttt{.OP} analysis:
\begingroup
\fontsize{8}{12}\selectfont
\begin{equation}
A_v^{\mathrm{OP}}\approx
-\frac{g_m^{\mathrm{OP}}}
{g_{ds}^{\mathrm{OP}}+1/R_L}.
\label{eq:op_anchored_gain}
\end{equation}
\endgroup
The three $g_m/I_D$ profiles establish different transistor bias conditions and inversion regions, producing corresponding variations in the extracted $g_m$ and $g_{ds}$. These parameters reflect the transistor characteristics at each operating point without requiring the long-channel approximations in Eq.~\eqref{eq:long_channel_params}. Consequently, discrepancies between the analytical prediction and the reference simulation more directly indicate limitations of the circuit-level relationship, such as neglected device contributions or loading effects. This allows the subsequent repair process to focus on refining the equation rather than approximating the underlying transistor characteristics.

\definecolor{iter1bg}{RGB}{255,245,243}
\definecolor{iter2bg}{RGB}{255,250,230}
\definecolor{iter3bg}{RGB}{241,251,244}

\definecolor{iter1frame}{RGB}{185,65,55}
\definecolor{iter2frame}{RGB}{185,135,20}
\definecolor{iter3frame}{RGB}{35,130,70}

\definecolor{errorredbg}{RGB}{255,220,220}
\definecolor{erroryellowbg}{RGB}{255,241,185}
\definecolor{errorgreenbg}{RGB}{218,245,225}

\definecolor{failred}{RGB}{185,35,35}
\definecolor{passgreen}{RGB}{25,125,65}
\definecolor{darkblue}{RGB}{20,65,120}

\begin{figure}[t]
\centering

{\scriptsize\bfseries Modeling UGF of 2S-OTA}

\vspace{-3pt}
{\scriptsize\bfseries\color{iter1frame}
Repair iteration 1
\hfill
Worst error: \textcolor{failred}{39.4\% -- FAIL}}
\vspace{1pt}

\begin{lstlisting}[
basicstyle=\ttfamily\fontsize{6.5}{5.5}\selectfont,
frame=single,
rulecolor=\color{iter1frame},
backgroundcolor=\color{iter1bg},
breaklines=true,
breakatwhitespace=true,
breakautoindent=false,
breakindent=0pt,
columns=fixed,
basewidth=0.55em,
keepspaces=true,
showstringspaces=false,
escapeinside={(*@}{@*)},
xleftmargin=0pt,
xrightmargin=0pt,
aboveskip=0pt,
belowskip=0pt
]
(*@\textbf{EQUATION:}@*)abs(((0.5*(gm_M1+gm_M0*gm_M3/(gm_M2+gds_M2+gds_M0)))/(gds_M1
+gds_M3))*(gm_M7/(gds_M6+gds_M7)))*((gds_M1+gds_M3)/(2*math.pi*(cdd_M1
+cdd_M3+cgg_M7+Cc*(1+abs(gm_M7/(gds_M6+gds_M7)))))))


(*@\textbf{EQUATION\_RATIONALE:}@*) UGF is the dominant-pole gain-BW product, using the same DC gain and the net3 pole used in the gain and BW equations.
\end{lstlisting}

\begin{lstlisting}[
basicstyle=\ttfamily\bfseries\fontsize{6.5}{7}\selectfont,
frame=single,
rulecolor=\color{iter1frame},
backgroundcolor=\color{errorredbg},
columns=fullflexible,
keepspaces=true,
showstringspaces=false,
xleftmargin=0pt,
xrightmargin=0pt,
aboveskip=0pt,
belowskip=0pt,
literate=
 {PASS}{{\textcolor{passgreen}{PASS}}}4
 {FAIL}{{\textcolor{failred}{FAIL}}}4
]
ERRORS: OP1 +19.4% FAIL | OP2 +31.1% FAIL | OP3 +39.4% FAIL
\end{lstlisting}

\centerline{\color{darkblue}\large$\Downarrow$}
\vspace{-10pt}

{\scriptsize\bfseries\color{iter2frame}
Repair iteration 3
\hfill
Worst error: \textcolor{failred}{5.1\% -- FAIL}}
\vspace{1pt}

\begin{lstlisting}[
basicstyle=\ttfamily\fontsize{6.5}{5.5}\selectfont,
frame=single,
rulecolor=\color{iter2frame},
backgroundcolor=\color{iter2bg},
breaklines=true,
breakatwhitespace=true,
breakautoindent=false,
breakindent=0pt,
columns=fixed,
basewidth=0.55em,
keepspaces=true,
showstringspaces=false,
escapeinside={(*@}{@*)},
xleftmargin=0pt,
xrightmargin=0pt,
aboveskip=0pt,
belowskip=0pt
]
(*@\textbf{REPAIR REASONING:}@*)
Step 1: The netlist is a two-stage Miller-compensated OTA: M0/M1 drive net1/net3 through a PMOS mirror load M2/M3, and M7/M6 form the second gain stage at vout. Step 2: The prior two-pole UGF solution collapsed to a very low frequency because it mixed Hz and rad/s style pole terms and over-weighted the compensation capacitor path. Step 3: Since the locked DC gain and locked 3dB bandwidth already pass, UGF should start from the dominant-pole identity (gain)*(3dB_BW), then include the physically visible Miller feed-forward degradation through Cc and the second-stage output capacitance. Step 4: I use the locked Av_total and locked Bandwidth_3dB_Hz product, reduced by a capacitive divider term Cc/(Cc + C_load + cdd_M6 + cdd_M7 + abs(cgd_M7)) weighted by the second-stage gm/gds loading path.

(*@\textbf{EQUATION:}@*)abs(((0.5*(gm_M1+gm_M0*gm_M3/(gm_M2+gds_M2+gds_M0))/(gds_M1+
gds_M3))*(gm_M7/(gds_M6+gds_M7)))*((gds_M1+gds_M3)/(2*math.pi*(cdd_M1+
cdd_M3+cgg_M7+Cc*(1+abs(gm_M7/(gds_M6+gds_M7)))+Cc*abs(gm_M7/(gds_M6+
gds_M7))*gm_M7/(gm_M6+gm_M7+gmb_M6+gmb_M7+abs(gm_M7/(gds_M6+gds_M7))*
gds_M6))))/(1+(Cc/(Cc+C_load+cdd_M6+cdd_M7+abs(cgd_M7)))*(gm_M7/(gm_M6+
gm_M7+gmb_M6+gmb_M7+abs(gm_M7/(gds_M6+gds_M7))*gds_M6)))))


(*@\textbf{EQUATION RATIONALE:}@*)
Uses the passing dominant-pole gain-bandwidth product and corrects the overestimate from second-stage Miller feed-forward loading through Cc into the finite vout capacitance.
\end{lstlisting}

\begin{lstlisting}[
basicstyle=\ttfamily\bfseries\fontsize{6.5}{7}\selectfont,
frame=single,
rulecolor=\color{iter2frame},
backgroundcolor=\color{erroryellowbg},
columns=fullflexible,
keepspaces=true,
showstringspaces=false,
xleftmargin=0pt,
xrightmargin=0pt,
aboveskip=0pt,
belowskip=0pt,
literate=
 {PASS}{{\textcolor{passgreen}{PASS}}}4
 {FAIL}{{\textcolor{failred}{FAIL}}}4
]
ERRORS: OP1 -5.1% FAIL | OP2 +1.4% PASS | OP3 +3.9% PASS
\end{lstlisting}

\centerline{\color{darkblue}\large$\Downarrow$}
\vspace{-10pt}

{\scriptsize\bfseries\color{iter3frame}
Repair iteration 5
\hfill
Worst error: \textcolor{passgreen}{3.4\% -- PASS}}
\vspace{1pt}

\begin{lstlisting}[
basicstyle=\ttfamily\fontsize{6.5}{5.5}\selectfont,
frame=single,
rulecolor=\color{iter3frame},
backgroundcolor=\color{iter3bg},
breaklines=true,
breakatwhitespace=true,
breakautoindent=false,
breakindent=0pt,
columns=fixed,
basewidth=0.55em,
keepspaces=true,
showstringspaces=false,
escapeinside={(*@}{@*)},
xleftmargin=0pt,
xrightmargin=0pt,
aboveskip=0pt,
belowskip=0pt
]
(*@\textbf{REPAIR REASONING:}@*)
Step 1: The UGF path is the locked first-stage transconductance at net3 multiplied by the second-stage M7/M6 gain and the locked dominant net3 pole. Step 2: The previous equation underestimated op_0 by 5.101%, while op_1 and op_2 already passed, so the repair must be a small correction to the compensation-return attenuation, not to DC gain or the dominant pole. Step 3: The missing physical term is the M6 drain conductance at vout in the local return denominator of the Cc-Rc feed-forward path; M6 is headroom-limited, and its finite gds damps the current returned through the compensation network. Step 4: I include +gds_M6 in that denominator only, which raises the low-bias UGF without changing the locked gain or bandwidth equations.

(*@\textbf{EQUATION:}@*)abs(((0.5*(gm_M1+gm_M0*gm_M3/(gm_M2+gds_M2+gds_M0))/(gds_M1+
gds_M3))*(gm_M7/(gds_M6+gds_M7)))*((gds_M1+gds_M3)/(2*math.pi*(cdd_M1+
cdd_M3+cgg_M7+Cc*(1+abs(gm_M7/(gds_M6+gds_M7)))+Cc*abs(gm_M7/(gds_M6+
gds_M7))*gm_M7/(gm_M6+gm_M7+gmb_M6+gmb_M7+abs(gm_M7/(gds_M6+gds_M7))*
gds_M6))))/(1+(Cc/(Cc+C_load+cdd_M6+cdd_M7+abs(cgd_M7)))*(gm_M7/(gm_M6+
gm_M7+gmb_M6+gmb_M7+abs(gm_M7/(gds_M6+gds_M7))*gds_M6+gds_M6)))))

(*@\textbf{EQUATION RATIONALE:}@*)
Locked passing UGF estimate from total low-frequency gain and corrected dominant-pole bandwidth.
\end{lstlisting}

\begin{lstlisting}[
basicstyle=\ttfamily\bfseries\fontsize{6.5}{7}\selectfont,
frame=single,
rulecolor=\color{iter3frame},
backgroundcolor=\color{errorgreenbg},
columns=fullflexible,
keepspaces=true,
showstringspaces=false,
xleftmargin=0pt,
xrightmargin=0pt,
aboveskip=0pt,
belowskip=0pt,
literate=
 {PASS}{{\textcolor{passgreen}{PASS}}}4
 {FAIL}{{\textcolor{failred}{FAIL}}}4
]
ERRORS: OP1 -3.2% PASS | OP2 +1.3% PASS | OP3 +3.4% PASS
\end{lstlisting}



\caption{Modeling UGF of the 2S-OTA testcase.}
\label{fig:ugf_repair_evolution_2s_ota}
\vspace{-5mm}
\end{figure}

\begin{figure*}[t]
    \centering
    \includegraphics[width=\textwidth]{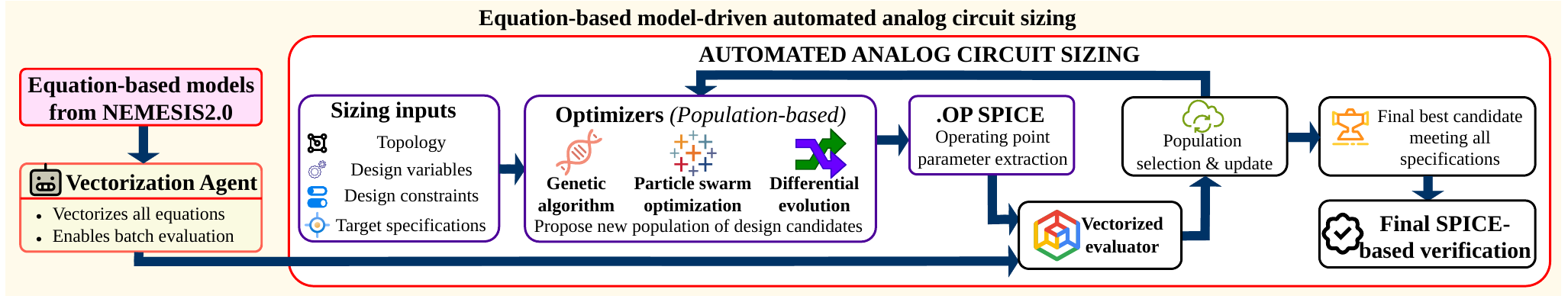}
    \vspace{-4mm}
    \caption{Overview of the proposed automated analog circuit sizing framework.}
    \label{fig:agentic_sizing}
    \vspace{-5mm}
\end{figure*}

For each metric $m$ and operating-point profile $k$, we evaluate the generated equation using the corresponding extracted device parameters to obtain the prediction $\widehat{y}_{m,k}$, while the metric-specific SPICE simulation provides the reference $y_{m,k}^{\mathrm{SPICE}}$. The percentage error is calculated as
\begingroup
\fontsize{8}{12}\selectfont
\begin{equation}
e_{m,k}=100\,
\frac{\left|\widehat{y}_{m,k}-y_{m,k}^{\mathrm{SPICE}}\right|}
{\max\!\left(\left|y_{m,k}^{\mathrm{SPICE}}\right|,\delta_m\right)}
\%,
\label{eq:verification_error}
\end{equation}
\endgroup
where $\delta_m>0$ is a metric-specific normalization floor that prevents excessive relative errors when the SPICE reference approaches zero. We accept an equation only if
\begin{equation}
e_{m,k}\leq\epsilon,\qquad
\forall\,k\in\{1,2,3\},
\label{eq:verification_criterion}
\end{equation}
where $\epsilon$ denotes the prescribed verification tolerance. By enforcing this criterion independently at every profile, we ensure that accuracy at one operating condition cannot compensate for inaccuracies at another. Equations that fail at any profile are forwarded to the \textit{Equation Repair Agent} with their predictions, reference values, and profile-specific errors. The extracted device parameters and reference values are retained for subsequent repair iterations, avoiding repeated simulations.

\subsubsection{\textbf{Equation repair}}
\label{sec:equation_repair}

\noindent
The \textit{Equation Repair Agent} revises analytical performance equations that fail the verification criterion (Eq.~\ref{eq:verification_criterion}) for one or more operating-point profiles. We provide the LLM with the circuit netlist, functional interpretation, candidate equations, and profile-specific verification feedback, prompting it to identify neglected device contributions, loading effects, compensation-network interactions, parasitic capacitances, and bias-dependent behavior. Each proposed revision must include a physical justification for any added, removed, or modified terms. We instruct the agent to avoid arbitrary fitting coefficients while permitting numerical factors derived from circuit relationships or unit conversions, such as $0.5$ for differential excitation or $2\pi$ for frequency conversion. We then evaluate each revised equation across all three profiles using previously extracted device parameters and reference values, without repeating simulations. Equations satisfying the prescribed error threshold at every profile are locked, while those failing at any profile are returned to the agent with updated errors and previous unsuccessful revisions. Repair therefore continues selectively for metrics that remain unverified.

Figure~\ref{fig:repair_prompt_context} illustrates the information provided to the \textit{Equation Repair Agent} for the 2S-OTA. The left panel presents the diagnostic instructions and constraints governing equation revisions. 
The upper-right panel uses unity-gain frequency (UGF) as a representative case to compare equation predictions against SPICE across the three operating-point profiles. Equations rejected in earlier repair iterations are also provided to the agent but omitted from the figure for brevity. The lower-right panel summarizes the verification errors and status of all performance metrics, including the initial DC\_gain, BW\_3dB, and OCMR equations presented in Section~\ref{sec:equation_generation}. While DC\_gain passes verification and is retained, BW\_3dB and OCMR require repair. The UGF errors in the detailed comparison also appear in the corresponding row of the overall summary, where a single status indicates whether the equation satisfies the verification criterion across all three profiles. 

Figure~\ref{fig:ugf_repair_evolution_2s_ota} shows how the agent uses this feedback to progressively refine the UGF equation, presenting iterations~1, 3, and 5 from the five-iteration repair history. In iteration~1, UGF is estimated using the low-frequency gain and dominant-pole frequency at \texttt{net3}, overestimating SPICE by 19.4\%, 31.1\%, and 39.4\% across the three profiles, consistent with the initial errors in Figure~\ref{fig:repair_prompt_context}. An intermediate two-pole formulation (not shown) substantially underestimates UGF, which the agent attributes to inconsistent frequency units (Hz versus rad/s) and excessive weighting of the compensation path. In iteration~3, the agent returns to the dominant-pole relationship $|A_{v0}|f_{3\mathrm{dB}}$, using the DC-gain and 3-dB-bandwidth equations verified by that iteration, and incorporates previously neglected feed-forward coupling through $C_c$, finite output-node capacitance, and second-stage $g_m/g_{ds}$ loading. The resulting errors are $-$5.1\%, 1.4\%, and 3.9\% at OP1, OP2, and OP3, respectively. Although OP2 and OP3 satisfy the verification criterion, OP1 remains outside the prescribed tolerance.

Based on the remaining discrepancy at OP1, the agent identifies an insufficient representation of the compensation-return path and introduces the finite drain conductance $g_{ds,M6}$ into the local return denominator of the $C_c$--$R_c$ feed-forward path. This correction accounts for the contribution of M6 to the output-node conductance while preserving the previously verified gain and bandwidth relationships. By iteration~5, the UGF errors decrease to $-$3.2\%, 1.3\%, and 3.4\% at OP1, OP2, and OP3, respectively, satisfying the verification criterion across all three profiles. The revised UGF equation is consequently locked. Upon successful verification of all requested performance metrics, we store the verified equations and their circuit context in the RAG knowledge base for future retrieval. The resulting equation set also serves as the reusable inner-loop performance evaluator in the automated sizing framework presented in the following section.

\vspace{-3mm}
\section{AUTOMATED SIZING FRAMEWORK}
\label{sec:agentic_sizing}

\noindent
Figure~\ref{fig:agentic_sizing} illustrates a framework for applying the SPICE-verified performance equations generated by \textsc{NEMESIS2.0} for schematic-level circuit sizing. For a given circuit topology, the independent sizing variables, design constraints, and target specifications define the optimization problem. The sizing algorithms are based on population-based stochastic search, where a significant number of candidates must be evaluated in each iteration.  To facilitate this, the equations are first taken through a \textit{Vectorization Agent} (Section~\ref{sec:vectorized_evaluation}) that transforms the verified scalar equations into numerically equivalent batch evaluators. 
For the circuit sizing problem (Section~\ref{sec:sizing_formulation}), NEMESIS2.0 models are used to evaluate constraints, rank candidates, and guide population updates \blue{(Section~\ref{sec:population_dse})}. The selected design is then verified against the original specifications using metric-specific SPICE testbenches~\cite{align_testbenches}: if any specification is violated, the corresponding internal optimization target is adjusted based on the observed discrepancy, and sizing is repeated \blue{(Section~\ref{sec:final_spice_validation}).}
\vspace{-2mm}

\subsection{Vectorized model generation}
\label{sec:vectorized_evaluation}

\noindent
Population-based circuit sizing requires evaluating the performance of multiple candidate designs at each optimization iteration. Although the verified scalar equations generated by \textsc{NEMESIS2.0} can evaluate individual candidates, executing them separately across a population introduces the overheads of repeated function calls to the model evaluator. To improve computational efficiency, we configure a \textit{Vectorization Agent} to transform these equations into batch evaluators using array-wise operations. The agent is instructed to preserve the original analytical expressions, metric definitions, numerical constants, safeguards, and operating-point-dependent conditions, modifying only their computational implementation.

For each performance metric $m$, the agent transforms the verified scalar function $f_{m,\mathcal{N}}$ defined in Eq.~\eqref{eq:eqgen_mapping} into a batch evaluator $F_{m,\mathcal{N}}$:
\begin{equation}
\label{eq:vectorized_estimator}
\widehat{\mathbf{y}}_m
=
F_{m,\mathcal{N}}
\!\left(
\{\mathbf{X}_{d}\}_{d\in\mathcal{D}},\mathbf{U}
\right),
\end{equation}
where $\mathbf{X}_{d}$ contains the extracted device parameters of transistor $d$ across the candidate batch and $\mathbf{U}$ contains the corresponding external circuit and testbench quantities. For a batch of $B$ candidates, $\widehat{\mathbf{y}}_m$ contains $B$ predicted values of metric $m$. Applying the evaluators across all requested metrics produces the performance matrix $\widehat{\mathbf{Y}}$, with candidates as rows and metrics as columns. Computational savings arise from array-wise equation evaluation rather than a reduction in the number of candidates or the operating-point analyses required to obtain their device parameters.

We verify the numerical consistency of each vectorized evaluator by comparing its predictions against those obtained from the corresponding verified scalar function using identical candidate inputs. Let $\widehat{\mathbf{y}}_{m}^{\mathrm{Vec}}$ denote the vectorized predictions and $\widehat{\mathbf{y}}_{m}^{\mathrm{Scalar}}$ the predictions obtained by evaluating the scalar function independently for each candidate. We require
$
\label{eq:vectorization_equivalence}
\left|
\widehat{\mathbf{y}}_{m}^{\mathrm{Vec}}
-
\widehat{\mathbf{y}}_{m}^{\mathrm{Scalar}}
\right|
\leq
\epsilon_{\mathrm{abs}}
+
\epsilon_{\mathrm{rel}}
\left|
\widehat{\mathbf{y}}_{m}^{\mathrm{Scalar}}
\right|,
\forall\,m,
$
where $\epsilon_{\mathrm{abs}}$ and $\epsilon_{\mathrm{rel}}$ are the prescribed absolute and relative numerical tolerances, respectively. The inequality is evaluated element-wise across the validation candidates. Any mismatch triggers correction of the vectorized implementation while the verified scalar equations remain unchanged. Evaluators satisfying this criterion on the validation inputs are accepted for circuit sizing.
\vspace{-2em}

\subsection{Circuit sizing problem formulation}
\label{sec:sizing_formulation}

\noindent
We formulate the circuit sizing problem independently of the optimization algorithm so that different optimizers can use the same design variables, constraints, and objective function. Each candidate proposed by the optimizer is represented by a vector of independent sizing variables:
\begingroup
\fontsize{8}{12}\selectfont
$$
\begin{aligned}
\boldsymbol{\theta}
=
[\theta_1,\theta_2,\ldots,\theta_{n_\theta}]^{\mathsf T},
\quad
\quad
\bar{\boldsymbol{\theta}}
=
\mathcal{Q}(\boldsymbol{\theta})\in\prod_{j=1}^{n_\theta}\Theta_j,
\end{aligned}
$$
\endgroup
where $n_\theta$ is the number of independent variables, which may include transistor widths, bias currents, and passive-component values. Each $\Theta_j$ defines the set of permitted values for variable $\theta_j$ based on its bounds and discretization. However, the optimizer-proposed values $\theta_j$ may not satisfy these requirements. Therefore, $\mathcal{Q}(\cdot)$ applies clipping and discretization to enforce the bounds and discretization requirements, producing the valid sizing vector $\bar{\boldsymbol{\theta}}$. 

For a batch of $B$ candidate designs, we instantiate each circuit using its valid sizing vector $\bar{\boldsymbol{\theta}}$. SPICE \texttt{.OP} analyses then provide the device parameters required by the performance equations. These parameters are organized into the arrays $\mathbf{X}_d$ and supplied, together with the external circuit and testbench parameters $\mathbf{U}$, to the vectorized evaluators in Eq.~\eqref{eq:vectorized_estimator}. The resulting performance matrix $\widehat{\mathbf{Y}}$ provides the predicted performance for all candidates.

The extracted operating-point parameters and predicted performance are then used to evaluate the corresponding constraint violations for each candidate. We first consider operating-point constraints, which impose biasing conditions on individual devices. For example, a transistor $d$ required to operate in saturation may be assigned the condition
$|V_{DS,d}|-|V_{\mathrm{DSAT},d}|\geq V_{\mathrm{margin}}^{\mathrm{sat}}$,
where $V_{\mathrm{margin}}^{\mathrm{sat}}$ is the prescribed voltage margin for saturation. 
For a given operating-point constraint $\ell$, we define the normalized violation vector $\mathbf{v}_{\ell}^{\mathrm{OP}}$ across candidates. For the saturation condition above, the violation for candidate $i$ is given by
$v_{\ell,i}^{\mathrm{OP}}=\max(0,V_{\mathrm{margin}}^{\mathrm{sat}}-|V_{DS,d,i}|+|V_{\mathrm{DSAT},d,i}|)/V_{DD}$,
where $V_{DD}=1\,\mathrm{V}$ is the supply voltage used for normalization. These violations are evaluated using the extracted operating-point parameters and are zero when the corresponding constraints are satisfied. Performance constraints, in contrast, are evaluated using the predicted metrics in $\widehat{\mathbf{Y}}$. For each metric $m$, we define the normalized constraint-violation vector as
\begingroup
\fontsize{8}{12}\selectfont
\begin{equation}
\label{eq:normalized_metric_violation}
\mathbf{v}_m=
\begin{cases}
\dfrac{\max(\mathbf{0},\widehat{\mathbf{y}}_m-U_m)}{s_m},
& \text{upper bound},\\[5pt]
\dfrac{\max(\mathbf{0},L_m-\widehat{\mathbf{y}}_m)}{s_m},
& \text{lower bound},\\[5pt]
\dfrac{\max(\mathbf{0},|\widehat{\mathbf{y}}_m-T_m|-\tau_m)}{s_m},
& \text{target},\\[5pt]
\dfrac{\max(\mathbf{0},L_m-\widehat{\mathbf{y}}_m,
\widehat{\mathbf{y}}_m-U_m)}{s_m},
& \text{interval},
\end{cases}
\end{equation}
\endgroup
where $L_m$ and $U_m$ are the lower and upper specification limits, $T_m$ is the target value, $\tau_m\geq 0$ is the designer-specified absolute tolerance defining the acceptable range $[T_m-\tau_m,T_m+\tau_m]$, with its specific values provided in Section~\ref{subsec:stage1_model_accuracy}, and $s_m>0$ is a metric-specific normalization scale determined from the corresponding lower and upper specification bounds. The operations are applied element-wise across candidates, producing zero violation when a specification is satisfied and a positive value otherwise. Operating-point constraint violations are normalized similarly using their corresponding electrical scales. 

The sizing formulation thus considers design-variable bounds enforced through $\mathcal{Q}(\cdot)$, and operating-point and performance constraints incorporated into the objective function through their normalized violations. We combine these violations with the design objectives into an objective vector:
\begingroup
\fontsize{8}{12}\selectfont
\begin{equation}
\label{eq:cost}
\mathbf{J}
=
\mathbf{J}_{\mathrm{soft}}
+
\rho
\sum_{m\in\mathcal{C}_{\mathrm{perf}}^{\mathrm{hard}}}
w_m(\mathbf{v}_m\odot\mathbf{v}_m)
+
\rho
\sum_{\ell\in\mathcal{C}_{\mathrm{OP}}^{\mathrm{hard}}}
\omega_\ell
(\mathbf{v}_{\ell}^{\mathrm{OP}}\odot\mathbf{v}_{\ell}^{\mathrm{OP}}),
\end{equation}
\endgroup
where $\mathbf{J}_{\mathrm{soft}}$ contains the configured design objectives, expressed as minimization terms. The sets $\mathcal{C}_{\mathrm{perf}}^{\mathrm{hard}}$ and $\mathcal{C}_{\mathrm{OP}}^{\mathrm{hard}}$ contain the mandatory performance and operating-point constraints, respectively. The coefficient $\rho>0$ controls the penalty contribution, while $w_m$ and $\omega_\ell$ are the corresponding nonnegative weights. 
\vspace{-1em}

\subsection{Population-based optimizers}
\label{sec:population_dse}

\begin{table}[t]
\centering
\caption{Circuit testcases and modeled performance metrics.}
\label{tab:circuit_benchmarks_metrics_single}

\scriptsize
\setlength{\tabcolsep}{2pt}
\renewcommand{\arraystretch}{1.12}

\begin{tabularx}{\columnwidth}{
>{\raggedright\arraybackslash}p{0.4\columnwidth}
>{\raggedright\arraybackslash}X}
\hline

\rowcolor{gray!18}
\rule{0pt}{3.0ex}
\textbf{Circuit class / Topologies} &
\textbf{Performance metrics} \\
\hline

\multirow[t]{5}{0.4\columnwidth}[-10pt]{%
\textbf{OTAs}\\
OTA-1: 5-transistor OTA\\
OTA-2: Current-mirror OTA\\
OTA-3: 2-stage OTA\\
OTA-4: Low-voltage cascode OTA\\
OTA-5: Telescopic OTA}
&
\textbf{AC:} DC gain (A$_{\mathrm{DC}}$), Unity-gain frequency (UGF),
3-dB bandwidth (BW$_{3\mathrm{dB}}$), Phase margin (PM),
Common-mode Rejection ratio (CMRR), Power-supply rejection ratio (PSRR+/-)
\\
&
\textbf{DC sweep:} Input and output common-mode range (ICMR$_{\mathrm{min/max}}$ and OCMR$_{\mathrm{min/max}}$)
\\
&
\textbf{Transient:} Slew rate (SR+/-)
\\
&
\textbf{Power:} DC power consumption (P$_{\mathrm{DC}}$)
\\
&
\textbf{Area:} Active-device area (A$_{\mathrm{dev}}$)
\\
\hline

\textbf{Comparator}\newline
COMP-1: StrongARM comparator\newline
COMP-2: Double-tail comparator
&
\textbf{Transient:} Evaluation-delay ($t_{\mathrm{eval}}$), Energy per decision ($E_{\mathrm{dec}}$), Precharge delay ($t_{\mathrm{pre}}$)\newline
\textbf{Transient sweep:} ICMR$_{\mathrm{min/max}}$\newline
\textbf{Area:} Active-device area (A$_{\mathrm{dev}}$)
\\
\hline
\hline

\end{tabularx}
\vspace{-5mm}
\end{table}

\noindent
We employ population-based optimizers to determine the sizing parameters of a given circuit topology. In iteration $t$, each optimizer maintains a population of $N_p$ candidate designs:
\begin{equation}
\mathcal{P}^{(t)}
=
\left\{
\boldsymbol{\theta}^{(t)}_1,
\boldsymbol{\theta}^{(t)}_2,
\ldots,
\boldsymbol{\theta}^{(t)}_{N_p}
\right\},
\end{equation}
where $\boldsymbol{\theta}^{(t)}_i$ contains the independent sizing variables of candidate $i$. We process each new or modified candidate using the parameter mapping and constraint-evaluation procedure described in Section~\ref{sec:sizing_formulation}, supplying the extracted device parameters to the vectorized performance equations. Unchanged candidates reuse their cached operating-point parameters, performance predictions, and objective values. We represent the resulting candidate evaluation as
\begin{equation}
\left(
J_i,\chi_i,\widehat{\mathbf{y}}_i,\mathbf{v}_i
\right)
=
\mathcal{E}\!\left(\boldsymbol{\theta}_i\right),
\end{equation}
where $\mathcal{E}(\cdot)$ denotes the candidate-evaluation process, $J_i$ is the objective value, $\chi_i$ is the hard-feasibility indicator, $\widehat{\mathbf{y}}_i$ contains the predicted performance metrics, and $\mathbf{v}_i$ contains the normalized constraint violations.

Each optimizer uses these evaluation results to update its candidate population according to its optimization strategy. The evaluation--update cycle continues until a prescribed stopping condition is reached, such as obtaining a satisfactory feasible design or exhausting the evaluation budget. We retain the best feasible candidate encountered, based on its objective value. If no candidate satisfies all mandatory constraints, the optimization run is considered unsuccessful.

We consider three population-based optimizers: GA~\cite{holland_75}, PSO~\cite{kennedy_95}, and DE~\cite{storn_97}. GA generates new sizing configurations through selection, crossover, and mutation; PSO updates candidate positions based on individual and swarm-best solutions; and DE constructs trial sizing vectors using scaled differences between population members. All three algorithms use the same circuit-evaluation procedure, sizing constraints, and objective function, enabling a consistent comparison of their performance under identical evaluation budgets.
\vspace{-3mm}

\subsection{Final SPICE-based verification}
\label{sec:final_spice_validation}

\noindent
Once the optimizer identifies a feasible design using the analytical performance model, we instantiate the optimized circuit and evaluate its performance using metric-specific SPICE testbenches~\cite{align_testbenches}. Unlike the earlier equation-verification stage, which assesses model accuracy across representative operating-point profiles, this final verification determines whether the optimized circuit satisfies the original application specifications. We compare the simulated performance with both the specifications and the model predictions to identify any discrepancies at the optimized operating point.

If any specification is violated, we tighten the corresponding internal optimization target based on the observed discrepancy and repeat the sizing process using the same verified vectorized equations. The resulting candidate is then reevaluated in SPICE. This verification--retuning cycle continues until all specifications are satisfied or the allotted optimization budget is exhausted. Only designs that meet the original specifications in the final SPICE simulations are considered successful. The SPICE-verified sizing results are presented in Section~\ref{sec:experiments_results}.

\section{Experiments and Results}
\label{sec:experiments_results}

\begin{table*}[t]
\centering
\caption{OTA model results: {\setlength{\fboxsep}{0pt}\colorbox{blue!8}{accuracy ($\mu$, $\sigma$, MAE)}}, {\setlength{\fboxsep}{0pt}\colorbox{green!10}{error tolerances}}, and {\setlength{\fboxsep}{0pt}\colorbox{orange!12}{convergence ($k_f$, $\varepsilon_f$)}}.}
\label{tab:stage1_ota_accuracy}

\setlength{\tabcolsep}{1pt}
\renewcommand{\arraystretch}{1}
\arrayrulecolor{black}

\resizebox{\textwidth}{!}{%
\begin{tabular}{
l|
>{\columncolor{blue!8}}c
>{\columncolor{blue!8}}c
>{\columncolor{blue!8}}c|
>{\columncolor{orange!12}}c
>{\columncolor{orange!12}}c|
>{\columncolor{blue!8}}c
>{\columncolor{blue!8}}c
>{\columncolor{blue!8}}c|
>{\columncolor{orange!12}}c
>{\columncolor{orange!12}}c|
>{\columncolor{blue!8}}c
>{\columncolor{blue!8}}c
>{\columncolor{blue!8}}c|
>{\columncolor{orange!12}}c
>{\columncolor{orange!12}}c|
>{\columncolor{blue!8}}c
>{\columncolor{blue!8}}c
>{\columncolor{blue!8}}c|
>{\columncolor{orange!12}}c
>{\columncolor{orange!12}}c|
>{\columncolor{blue!8}}c
>{\columncolor{blue!8}}c
>{\columncolor{blue!8}}c|
>{\columncolor{orange!12}}c
>{\columncolor{orange!12}}c|
>{\columncolor{green!10}}c
}
\hline

\rowcolor{gray!18}
&
\multicolumn{5}{c|}{\textbf{OTA-1}} &
\multicolumn{5}{c|}{\textbf{OTA-2}} &
\multicolumn{5}{c|}{\textbf{OTA-3}} &
\multicolumn{5}{c|}{\textbf{OTA-4}} &
\multicolumn{5}{c|}{\textbf{OTA-5}} &
\cellcolor{green!20}
\\

\rowcolor{gray!18}
\multirow{-2}{*}{\textbf{Metric}} &
\cellcolor{blue!18}$\mu$ &
\cellcolor{blue!18}$\sigma$ &
\cellcolor{blue!18}MAE &
\cellcolor{orange!22}$k_f$ &
\cellcolor{orange!22}$\varepsilon_f$ &
\cellcolor{blue!18}$\mu$ &
\cellcolor{blue!18}$\sigma$ &
\cellcolor{blue!18}MAE &
\cellcolor{orange!22}$k_f$ &
\cellcolor{orange!22}$\varepsilon_f$ &
\cellcolor{blue!18}$\mu$ &
\cellcolor{blue!18}$\sigma$ &
\cellcolor{blue!18}MAE &
\cellcolor{orange!22}$k_f$ &
\cellcolor{orange!22}$\varepsilon_f$ &
\cellcolor{blue!18}$\mu$ &
\cellcolor{blue!18}$\sigma$ &
\cellcolor{blue!18}MAE &
\cellcolor{orange!22}$k_f$ &
\cellcolor{orange!22}$\varepsilon_f$ &
\cellcolor{blue!18}$\mu$ &
\cellcolor{blue!18}$\sigma$ &
\cellcolor{blue!18}MAE &
\cellcolor{orange!22}$k_f$ &
\cellcolor{orange!22}$\varepsilon_f$ &
\cellcolor{green!20}\multirow{-2}{*}{\textbf{Tolerance}}
\\
\hline

A$_{\mathrm{DC}}$ (dB)
& -0.23 & 0.08 & 0.23 & 1 & 2.07
& 0.60 & 0.13 & 0.60 & 1 & 4.31
& -0.29 & 0.13 & 0.29 & 1 & 0.10
& 0.87 & 1.07 & 0.87 & 3 & 1.41
& 0.30 & 2.30 & 1.51 & 2 & 0.87
& Abs., 2 dB
\\

UGF (MHz)
& 0.33 & 0.12 & 0.33 & 1 & 1.49
& 0.16 & 0.26 & 0.21 & 8 & 3.23
& 2.01 & 4.75 & 3.10 & 5 & 3.42
& 0.23 & 1.07 & 0.25 & 3 & 3.94
& 0.07 & 1.23 & 0.81 & 2 & 4.37
& Rel., 5\%
\\

BW$_{3\mathrm{dB}}$ (MHz)
& 0.00 & 0.01 & 0.01 & 3 & 1.75
& 0.02 & 0.01 & 0.02 & 8 & 1.76
& 0.16 & 0.81 & 0.21 & 6 & 4.97
& 0.23 & 1.07 & 0.25 & 3 & 0.90
& 0.99 & 4.03 & 1.05 & 4 & 3.26
& Rel., 5\%
\\

PM ($^\circ$)
& -0.05 & 0.08 & 0.07 & 1 & 0.19
& 0.07 & 0.26 & 0.21 & 2 & 0.95
& -4.21 & 4.68 & 4.60 & 6 & 4.12
& -5.44 & 4.58 & 5.44 & 1 & 4.15
& -3.25 & 9.87 & 5.27 & 2 & 2.27
& Abs., 5$^\circ$
\\

CMRR (dB)
& 1.09 & 0.56 & 1.09 & 5 & 4.90
& 0.68 & 1.30 & 1.17 & 2 & 2.74
& -2.29 & 2.64 & 2.73 & 6 & 2.91
& 1.09 & 4.02 & 2.84 & 4 & 2.23
& -2.45 & 4.52 & 3.98 & 3 & 3.98
& Abs., 5 dB
\\

PSRR$+$ (dB)
& -0.12 & 0.06 & 0.12 & 2 & 1.28
& 0.40 & 1.08 & 0.91 & 2 & 4.84
& -1.15 & 1.46 & 1.59 & 4 & 4.94
& 1.36 & 2.70 & 2.27 & 4 & 4.03
& -2.02 & 2.16 & 2.87 & 2 & 4.26
& Abs., 2 dB
\\

PSRR$-$ (dB)
& -0.76 & 0.98 & 1.03 & 2 & 2.80
& -0.11 & 0.27 & 0.21 & 2 & 2.02
& -2.04 & 5.89 & 4.06 & 3 & 4.41
& 2.47 & 5.30 & 5.75 & 4 & 4.36
& -2.05 & 6.49 & 6.31 & 3 & 4.75
& Abs., 2 dB
\\

ICMR$_{\min}$ (mV)
& 20.74 & 35.42 & 32.35 & 4 & 3.36
& 1.86 & 33.13 & 26.28 & 4 & 4.78
& 12.68 & 17.24 & 16.36 & 5 & 4.46
& 33.70 & 30.31 & 38.02 & 3 & 4.86
& 25.34 & 35.65 & 34.16 & 3 & 4.62
& Abs., 50 mV
\\

ICMR$_{\max}$ (mV)
& 38.90 & 23.48 & 39.20 & 1 & 4.68
& 2.79 & 46.68 & 34.74 & 4 & 4.89
& -31.50 & 45.72 & 37.37 & 6 & 2.90
& -27.91 & 61.22 & 46.77 & 3 & 4.24
& 1.95 & 37.05 & 31.58 & 2 & 4.78
& Abs., 50 mV
\\

OCMR$_{\min}$ (mV)
& 15.60 & 21.15 & 20.13 & 5 & 4.08
& -0.72 & 3.76 & 2.45 & 4 & 4.64
& -25.77 & 143.40 & 33.27 & 13 & 4.91
& 20.47 & 18.90 & 22.18 & 4 & 3.48
& 17.40 & 22.76 & 20.22 & 2 & 2.36
& Abs., 50 mV
\\

OCMR$_{\max}$ (mV)
& -29.09 & 24.36 & 33.47 & 2 & 3.03
& 6.31 & 24.94 & 21.17 & 2 & 2.91
& -44.10 & 53.66 & 46.94 & 5 & 4.67
& 22.14 & 37.46 & 33.01 & 1 & 2.83
& -25.26 & 36.56 & 34.88 & 2 & 4.48
& Abs., 50 mV
\\

SR$+$ (V/$\mu$s)
& 1.42 & 3.48 & 2.66 & 5 & 4.96
& 0.93 & 2.51 & 1.68 & 6 & 2.37
& 5.35 & 8.95 & 7.05 & 11 & 4.77
& 0.80 & 0.73 & 0.81 & 4 & 2.70
& 1.37 & 2.31 & 1.66 & 3 & 3.50
& Rel., 10\%
\\

SR$-$ (V/$\mu$s)
& 2.08 & 3.24 & 2.70 & 6 & 4.56
& 0.84 & 2.54 & 1.78 & 6 & 3.93
& 5.29 & 4.83 & 5.84 & 13 & 4.92
& 2.33 & 2.14 & 2.36 & 4 & 4.90
& 0.07 & 0.90 & 0.70 & 3 & 1.80
& Rel., 10\%
\\

\hline
\end{tabular}%
}
\vspace{-3mm}
\end{table*}

\subsection{Experimental setup}
\label{sec:experimental_setup}

\noindent
We evaluate \textsc{NEMESIS2.0} and its automated sizing framework using seven circuit testcases comprising five OTAs and two COMPs. Table~\ref{tab:circuit_benchmarks_metrics_single} summarizes the testcases and their modeled performance metrics. All circuits are implemented in a commercial 65\,nm CMOS technology with a 1\,V supply and simulated using Synopsys HSPICE. To examine model adaptability to external loading conditions, we also evaluate the generated equations across output load capacitances ($C_L$) ranging from 100\,fF to 1\,pF.

All LLM-based agents use OpenAI GPT-5.5~\cite{gpt55} with high reasoning effort. We repeat equation generation and repair over five independent runs per circuit and report averaged results unless otherwise specified. The equation-verification tolerance in Eq.~\eqref{eq:verification_criterion} is set to $\epsilon=5\%$ for all modeled metrics. Our experiments assess the accuracy and convergence of the generated equations and their effectiveness in automated sizing under different performance targets. For comparisons across GA, PSO, and DE, we maintain identical sizing variables, bounds, constraints, initialization, and candidate-evaluation budgets. The open-source repository is available at \url{https://github.com/UMN-EDA/NEMESIS}.

\vspace{-3mm}

\subsection{Performance model accuracy and convergence}
\label{subsec:stage1_model_accuracy}

\noindent
We evaluate the quality of the generated performance equations across the seven circuit testcases from three complementary perspectives. First, we assess model generalization by comparing equation predictions against SPICE references at previously unseen sizing configurations and quantifying the resulting prediction errors. Second, we examine these errors relative to metric-specific tolerances, allowing accuracy to be compared across performance metrics with different units and numerical scales. Third, we evaluate the modeling effort by examining the number of iterations required to obtain verified equations and their final verification errors.

\begin{table}[t]
\centering
\caption{Comparator model results: {\setlength{\fboxsep}{0pt}\colorbox{blue!8}{accuracy ($\mu$, $\sigma$, MAE)}}, {\setlength{\fboxsep}{0pt}\colorbox{green!10}{error tolerances}}, and {\setlength{\fboxsep}{0pt}\colorbox{orange!12}{convergence ($k_f$, $\varepsilon_f$)}}.}
\label{tab:stage1_comp_accuracy}

\setlength{\tabcolsep}{1pt}
\renewcommand{\arraystretch}{1}
\arrayrulecolor{black}

\resizebox{\columnwidth}{!}{%
\begin{tabular}{
l|
>{\columncolor{blue!8}}c
>{\columncolor{blue!8}}c
>{\columncolor{blue!8}}c|
>{\columncolor{orange!12}}c
>{\columncolor{orange!12}}c|
>{\columncolor{blue!8}}c
>{\columncolor{blue!8}}c
>{\columncolor{blue!8}}c|
>{\columncolor{orange!12}}c
>{\columncolor{orange!12}}c|
>{\columncolor{green!10}}c
}
\hline
\rowcolor{gray!18}
&
\multicolumn{5}{c|}{\textbf{COMP-1}} &
\multicolumn{5}{c|}{\textbf{COMP-2}} &
\cellcolor{green!20}
\\

\rowcolor{gray!18}
\multirow{-2}{*}{\textbf{Metric}} &
\cellcolor{blue!18}$\mu$ &
\cellcolor{blue!18}$\sigma$ &
\cellcolor{blue!18}MAE &
\cellcolor{orange!22}$k_f$ &
\cellcolor{orange!22}$\varepsilon_f$ (\%) &
\cellcolor{blue!18}$\mu$ &
\cellcolor{blue!18}$\sigma$ &
\cellcolor{blue!18}MAE &
\cellcolor{orange!22}$k_f$ &
\cellcolor{orange!22}$\varepsilon_f$ (\%) &
\cellcolor{green!20}\multirow{-2}{*}{\textbf{Tolerance}}
\\
\hline

$t_{\mathrm{eval}}$ (ps)
& -10.67 & 17.80 & 16.10 & 12 & 4.85
& 11.80 & 13.00 & 15.00 & 6 & 3.72
& Rel., 10\%
\\

$E_{\mathrm{dec}}$ (fJ)
& 14.22 & 6.47 & 15.57 & 31 & 3.23
& 42.04 & 27.78 & 42.14 & 9 & 1.83
& Rel., 10\%
\\

ICMR$_{\min}$ (mV)
& 5.06 & 6.15 & 5.09 & 2 & 4.31
& -4.08 & 26.38 & 16.13 & 2 & 4.77
& Abs., 50 mV
\\

ICMR$_{\max}$ (mV)
& -5.12 & 20.93 & 10.87 & 30 & 1.11
& 3.32 & 0.47 & 3.32 & 2 & 0.39
& Abs., 50 mV
\\

$t_{\mathrm{pre}}$ (ps)
& -10.63 & 31.51 & 24.69 & 7 & 4.02
& 10.79 & 36.00 & 11.30 & 9 & 4.62
& Rel., 10\%
\\

\hline
\end{tabular}%
}
\vspace{-5mm}
\end{table}

\subsubsection{\textbf{Model generalization and error statistics}}

\noindent
To evaluate model generalization, we examine the predictive accuracy of the verified equations beyond the three representative operating-point profiles used during model construction. For each testcase, we generate an additional 1000 fully sized circuit candidates using Latin hypercube sampling (LHS)~\cite{santner2019design}, covering a wider range of transistor dimensions than those used during equation construction. These candidates are excluded from equation generation and repair, allowing us to assess the fixed equations at previously unseen operating points. We compare the predicted performance metrics against their corresponding SPICE references for each candidate.

For performance metric $m$ and LHS candidate $i$, the signed model--SPICE residual is defined as
$$
\Delta_{m,i}
=
y_{m,i}^{\mathrm{Model}}
-
y_{m,i}^{\mathrm{SPICE}},
$$
where $y_{m,i}^{\mathrm{Model}}$ and $y_{m,i}^{\mathrm{SPICE}}$ denote the model prediction and SPICE reference, respectively. Positive residuals indicate overestimation, while negative residuals indicate underestimation. The blue-shaded columns in Tables~\ref{tab:stage1_ota_accuracy} and~\ref{tab:stage1_comp_accuracy} summarize the prediction errors for the OTA and comparator testcases, respectively, using the following statistics:
\begingroup
\fontsize{8}{12}\selectfont
$$
\begin{gathered}
\mu_m=
\frac{1}{N_m}\sum_{i=1}^{N_m}\Delta_{m,i},
\qquad
\sigma_m=
\sqrt{\frac{1}{N_m-1}
\sum_{i=1}^{N_m}(\Delta_{m,i}-\mu_m)^2},
\\[-1pt]
\mathrm{MAE}_m=
\frac{1}{N_m}\sum_{i=1}^{N_m}|\Delta_{m,i}|.
\end{gathered}
$$
\endgroup
Here, $N_m$ denotes the number of evaluated LHS candidates for metric $m$. The mean residual ($\mu_m$) indicates systematic overestimation or underestimation, while the standard deviation ($\sigma_m$) measures the spread of prediction errors. The mean absolute error (MAE) quantifies the average error magnitude regardless of its sign. All three statistics are expressed in the units of the corresponding performance metric.

The results illustrate differences in prediction accuracy across metrics and circuit topologies. The signed mean residual complements MAE by identifying systematic overestimation or underestimation, which cannot be determined from absolute errors alone. For example, the DC-gain model of OTA-3 exhibits $\mu=-0.29$\,dB and MAE$=0.29$\,dB, indicating predominantly negative residuals and hence underestimation. In contrast, the SR$+$ model of OTA-2 has $\mu=0.93$\,V/$\mu$s, $\sigma=2.51$\,V/$\mu$s, and MAE$=1.68$\,V/$\mu$s. The difference between the signed mean and MAE reflects partial cancellation of positive and negative residuals, while the positive mean indicates an overall tendency toward overestimation.

\subsubsection{\textbf{Normalized model vs. SPICE error analysis}}

\begin{figure}[t]
    \centering
    \includegraphics[width=\columnwidth]{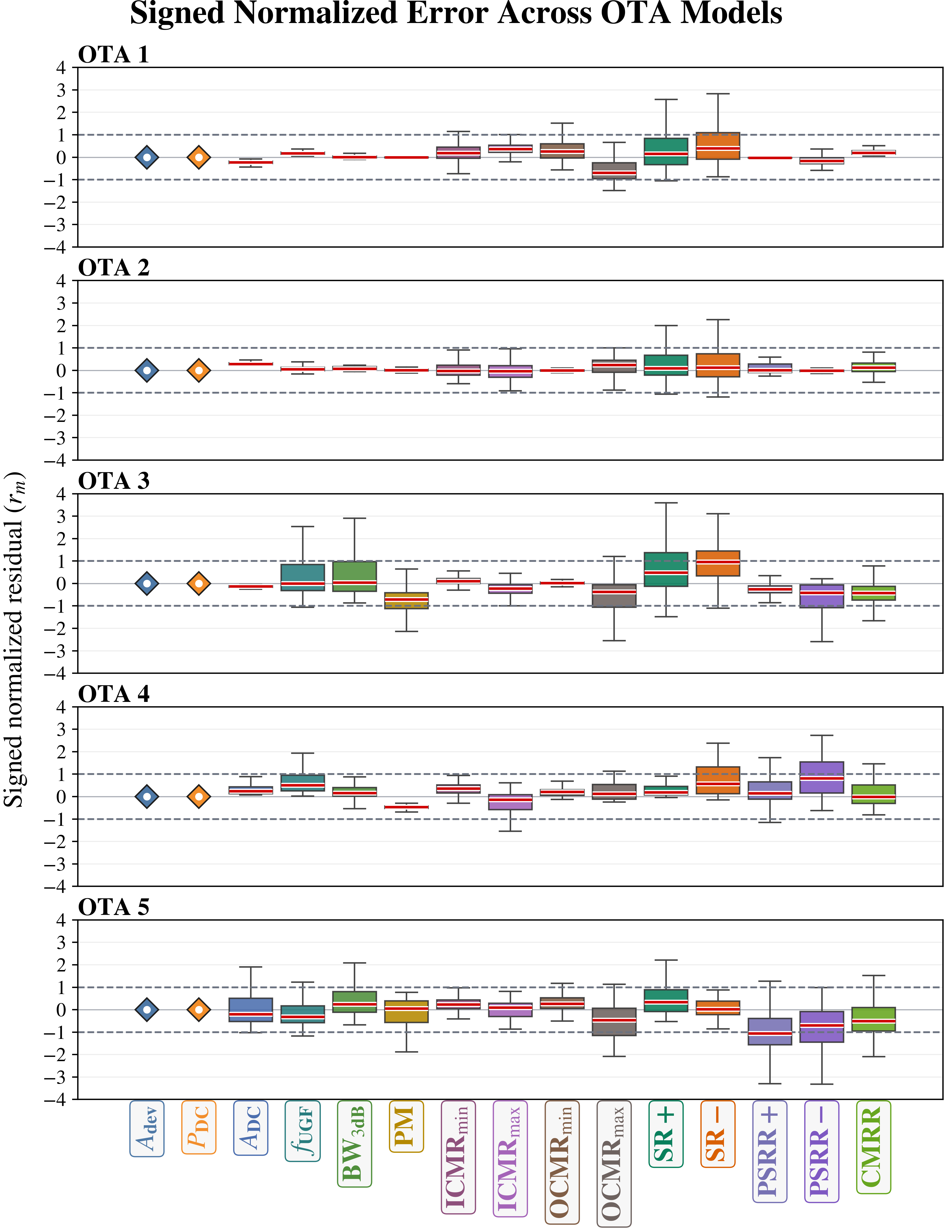}
    \caption{Distribution of the OTA model--SPICE residuals relative to their metric-specific error tolerances.}
    \label{fig:ota_boxplots_combined}
    \vspace{-8mm}
\end{figure}

\noindent
The preceding error statistics quantify prediction accuracy in the units of individual performance metrics, making direct comparisons across metrics difficult. To establish a common scale, we normalize each signed model--SPICE residual by its metric-specific error tolerance:
$$
r_{m,i} = \frac{\Delta_{m,i}}{\eta_{m,i}},
\quad
\eta_{m,i} =
\begin{cases}
\tau_m, & \text{absolute tolerance},\\
\rho_m |y_{m,i}^{\mathrm{SPICE}}|, & \text{relative tolerance}.
\end{cases}
$$
where $\eta_{m,i}$ is the metric-specific error-normalization scale, $\tau_m$ is the absolute tolerance, and $\rho_m$ is the relative tolerance factor.
The green-shaded columns in Tables~\ref{tab:stage1_ota_accuracy} and~\ref{tab:stage1_comp_accuracy} specify these tolerances for each metric. Absolute tolerances are used for $A_{\mathrm{DC}}$, PM, CMRR, PSRR$+/-$, ICMR$_{\min/\max}$, and OCMR$_{\min/\max}$, since their prediction errors are more meaningfully interpreted as differences in dB, degrees, or voltage. Relative tolerances are used for UGF, BW$_{3\mathrm{dB}}$, SR$+/-$, $E_{\mathrm{dec}}$, $t_{\mathrm{eval}}$, and $t_{\mathrm{pre}}$, for which fractional deviations from the reference values provide a meaningful measure of error. These tolerances represent acceptable prediction errors for circuit sizing and are distinct from the $5\%$ equation-verification threshold used during model construction. Accordingly, $r_{m,i}=0$ indicates exact agreement, $|r_{m,i}|\leq1$ indicates an error within tolerance, and $|r_{m,i}|>1$ indicates a tolerance exceedance.

Figure~\ref{fig:ota_boxplots_combined} presents the normalized residual distributions for the OTA metrics. The solid line at $r_{m,i}=0$ indicates exact model--SPICE agreement, the dashed lines at $\pm1$ mark the tolerance limits, and the thick red lines within the boxes represent mean residuals. Most distributions are centered near zero, although SR$+/-$ and PSRR$+/-$ exhibit comparatively larger variations. Here, $A_{\mathrm{dev}}$ and $P_{\mathrm{DC}}$ have zero residual across the evaluated candidates and are represented by diamond markers at $r_{m,i}=0$ and $A_{\mathrm{dev}}=\sum_{d\in\mathcal{D}}W_dL_d$ provides an area estimate at the schematic stage. 

\subsubsection{\textbf{Model convergence}}

\noindent
The orange-shaded columns in Tables~\ref{tab:stage1_ota_accuracy} and~\ref{tab:stage1_comp_accuracy} summarize equation-repair convergence. For each metric, $k_f$ denotes the repair iteration attaining the lowest recorded maximum verification error across the three operating-point profiles, and $\varepsilon_f$ is the corresponding error. All OTA and COMP equations satisfy $\varepsilon_f<5\%$, confirming successful verification under the prescribed criterion. However, the refinement effort varies across circuits and metrics. For OTA-3, OCMR$_{\min}$ and SR$-$ reach $k_f=13$, whereas all COMP-2 metrics converge within nine iterations. COMP-1 requires more extensive refinement for selected metrics, reaching $k_f=31$ for $E_{\mathrm{dec}}$ and $k_f=30$ for ICMR$_{\max}$. These results show that convergence behavior depends on the circuit topology and performance metric, even under a common verification criterion.
\vspace{-3mm}

\subsection{Model-driven sizing evaluation}
\label{subsec:stage2_evaluation}

\noindent
We evaluate the generated performance equations for automated sizing across seven testcases through three experiments. First, we perform application-specific sizing to examine performance tradeoffs under different optimization objectives. Second, we assess single-sizing-attempt feasibility across 100 specification sets per testcase and evaluate the effectiveness of target tightening in recovering from unsuccessful attempts. Third, we compare GA, PSO, and DE using the same equations to assess optimizer portability and quantify runtime savings relative to SPICE-in-the-loop sizing. We independently verify the resulting designs against their original specifications.

\begin{table}[t]
\centering
\caption{GA-based application-specific OTA sizing.}
\label{tab:ga_ota_sizing}

\scriptsize
\setlength{\tabcolsep}{3pt}
\renewcommand{\arraystretch}{0.90}
\arrayrulecolor{black}

\resizebox{\columnwidth}{!}{%
\begin{tabular}{
>{\raggedright\arraybackslash}p{1.8cm}|
c|c||
c|c||
c|c||
c|c}
\hline


\rowcolor{gray!18}
&
\multicolumn{2}{c||}{\textbf{High speed}} &
\multicolumn{2}{c||}{\textbf{Low power}} &
\multicolumn{2}{c||}{\textbf{Balanced}} &
\multicolumn{2}{c}{\textbf{High gain}}
\\

\hhline{
>{\arrayrulecolor{gray!18}}-
>{\arrayrulecolor{black}}|
--||--||--||--
}

\rowcolor{gray!18}
\multirow{-2}{*}{\textbf{Metric}} &
\textbf{Target} & \textbf{SPICE} &
\textbf{Target} & \textbf{SPICE} &
\textbf{Target} & \textbf{SPICE} &
\textbf{Target} & \textbf{SPICE}
\\
\arrayrulecolor{black}
\hline


\rowcolor{gray!10}
\multicolumn{9}{c}{\textbf{OTA-1 (5T-OTA)}}\\
\hline

A$_{\mathrm{DC}}$ (dB)         & $\ge$20 & 21.73 & $\ge$20 & 21.05 & $\ge$22 & 22.68 & $\ge$27 & 28.44 \\
UGF (MHz)                       & $\ge$600 & 683.66 & max. & 3.21 & $\ge$80 & 88.54 & max. & 6.70 \\
BW$_{3\mathrm{dB}}$ (MHz)      & $\ge$50 & 55.13 & max. & 0.29 & $\ge$5 & 6.05 & max. & 0.28 \\
PM ($^\circ$)                   & $\ge$60 & 88.76 & $\ge$60 & 83.14 &$\ge$70& 85.05 & $\ge$70 & 80.25 \\
CMRR (dB)                       & $\ge$45 & 48.79 & max. & 48.43 & $\ge$45 & 48.24 & $\ge$55 & 60.56 \\
PSRR$+$ (dB)                 & max. & 20.91 & max. & 21.89 & max. & 22.46 & $\ge$25 & 28.34 \\
PSRR$-$ (dB)                 & max. & 46.20 & max. & 52.28 & max. & 52.71 & $\ge$58 & 60.56 \\
ICMR$_{\min}$ (V)               & $\le$0.4& 0.39 & $\le$0.4 & 0.39 & $\le$0.4 & 0.40 & $\le$0.4 & 0.35 \\
ICMR$_{\max}$ (V)               & $\ge$0.8 & 0.87 & $\ge$0.8 & 0.88 & $\ge$0.8 & 0.85 & $\ge$0.8 & 0.85 \\
OCMR$_{\min}$ (V)               &  $\le$0.3 & 0.22 & $\le$0.2 & 0.18 & $\le$0.2 & 0.17 & $\le$0.2 & 0.19 \\
OCMR$_{\max}$ (V)               &  $\ge$0.8 & 0.82 & $\ge$0.8 & 0.85 & $\ge$0.8 & 0.85 & $\ge$0.8 & 0.86 \\
SR$+$ (V/$\mu$s)             &  $\ge$300 & 355.81 & max. & 12.57 & $\ge$40 & 47.29 & max. & 3.85 \\
SR$-$ (V/$\mu$s)             &  $\ge$300 & 379.12 & max. & 11.42 & $\ge$40 & 41.01 & max. & 5.28\\
P$_{\mathrm{DC}}$ ($\mu$W)      & min. & 523.55 & $\le$10 & 9.73 & min. & 42.97 & min. & 29.26 \\
A$_{\mathrm{dev}}$ ($\mu$m$^2$) & min. & 34.56 & min. & 24.08 & min. & 49.76 & min. & 232.08 \\

$t_f$ (sec) & \multicolumn{2}{c||}{13.58} & \multicolumn{2}{c||}{19.52} & \multicolumn{2}{c||}{18.04} & \multicolumn{2}{c}{15.26} \\

\hline

\rowcolor{gray!10}
\multicolumn{9}{c}{\textbf{OTA-2 (CM-OTA)}}\\
\hline
A$_{\mathrm{DC}}$ (dB)         & $\ge$20 & 20.56 & $\ge$20 & 23.54 & $\ge$22 & 22.60 & $\ge$30 & 30.54 \\
UGF (MHz)                       & $\ge$490 & 511.49 & max. & 26.03 & $\ge$80 & 138.00 & max. & 1.07 \\
BW$_{3\mathrm{dB}}$ (MHz)       & $\ge$50 & 51.75 & max. & 1.73 & $\ge$5 & 10.23 & max. & 0.031 \\
PM ($^\circ$)                   & $\ge$60 & 93.32 & $\ge$60 & 94.05 & $\ge$70 & 94.69 & $\ge$70 & 93.81 \\
CMRR (dB)                       & $\ge$45 & 51.62 & max. & 53.54 & $\ge$45 & 50.31 & $\ge$55 & 58.04 \\
PSRR$+$ (dB)                 & max. & 29.12 & max. & 58.01 & max. & 57.08 & $\ge$50 & 55.76 \\
PSRR$-$ (dB)                 & max. & 33.24 & max. & 25.50 & max. & 24.89 & $\ge$35 & 39.34 \\
ICMR$_{\min}$ (V)               & $\le$0.5 & 0.43 & $\le$0.3 & 0.26 & $\le$0.3 & 0.27 & $\le$0.3 & 0.28 \\
ICMR$_{\max}$ (V)               & $\ge$0.8 & 0.84 & $\ge$0.7 & 0.75 & $\ge$0.8 & 0.87 & $\ge$0.8 & 0.84 \\
OCMR$_{\min}$ (V)               & $\le$0.2 & 0.17 & $\le$0.2 & 0.15 & $\le$0.2 & 0.19 & $\le$0.3 & 0.24 \\
OCMR$_{\max}$ (V)               & $\ge$0.8 & 0.88 & $\ge$0.7 & 0.88 & $\ge$0.8 & 0.89 & $\ge$0.8 & 0.86 \\
SR$+$ (V/$\mu$s)             & $\ge$800 & 874.27 & max. & 11.49 & $\ge$20 & 41.08 & max. & 86.96 \\
SR$-$ (V/$\mu$s)             & $\ge$800 & 809.04 & max. & 12.45 & $\ge$30 & 55.85 & max. & 86.97 \\
P$_{\mathrm{DC}}$ ($\mu$W)      & min. & 1543.79 & $\le$30 & 28.42 & min. & 94.89 & min. & 364.80 \\
A$_{\mathrm{dev}}$ ($\mu$m$^2$) & min. & 15.24 & min. & 55.00 & min. & 55.40 & min. & 290.24 \\
$t_f$ (sec)                       & \multicolumn{2}{c||}{22.73} & \multicolumn{2}{c||}{21.44} & \multicolumn{2}{c||}{14.09} & \multicolumn{2}{c}{27.29} \\

\hline


\rowcolor{gray!10}
\multicolumn{9}{c}{\textbf{OTA-3 (2S-OTA)}}\\
\hline
A$_{\mathrm{DC}}$ (dB)         & $\ge$25 & 27.04 & $\ge$25 & 29.07 & $\ge$40 & 44.92 & $\ge$60 & 62.72 \\
UGF (MHz)                       & $\ge$500 & 582.97 & max. & 28.40 & $\ge$80 & 108.58 & max. & 42.90 \\
BW$_{3\mathrm{dB}}$ (MHz)       & $\ge$20 & 25.03 & max. & 1.39 & $\ge$1 & 1.08 & max. & 0.08 \\
PM ($^\circ$)                   & $\ge$60 & 89.19 & $\ge$60 & 70.60 & $\ge$70 & 82.95 & $\ge$60 & 65.87 \\
CMRR (dB)                       & $\ge$45 & 49.62 & $\ge$45 & 55.27 & $\ge$45 & 60.09 & $\ge$55 & 57.04 \\
PSRR$+$ (dB)                 & max. & 21.26 & max. & 34.75 & max. & 37.38 & $\ge$40 & 43.48 \\
PSRR$-$ (dB)                 & max. & 54.68 & max. & 34.80 & max. & 50.14 & $\ge$50 & 59.05 \\
ICMR$_{\min}$ (V)               & $\le$0.4 & 0.38 & $\le$0.4 & 0.33 & $\le$0.3 & 0.28 & $\le$0.3 & 0.27 \\
ICMR$_{\max}$ (V)               & $\ge$0.7 & 0.74 & $\ge$0.8 & 0.88 & $\ge$0.8 & 0.88 & $\ge$0.8 & 0.92 \\
OCMR$_{\min}$ (V)               & $\le$0.2 & 0.12 & $\le$0.2 & 0.17 & $\le$0.2 & 0.19 & $\le$0.3 & 0.28 \\
OCMR$_{\max}$ (V)               & $\ge$0.7 & 0.79 & $\ge$0.8 & 0.82 & $\ge$0.8 & 0.83 & $\ge$0.8& 0.89 \\
SR$+$ (V/$\mu$s)             & $\ge$400 & 488.16 & max. & 7.03 & $\ge$30 & 36.13 & max. & 4.63 \\
SR$-$ (V/$\mu$s)             & $\ge$400 & 475.76 & max. & 4.00 & $\ge$30 & 34.56 & max. & 2.94 \\
P$_{\mathrm{DC}}$ ($\mu$W)      & min. & 1163.66 & $\le$20 & 18.55 & min. & 86.92 & min. & 113.27 \\
A$_{\mathrm{dev}}$ ($\mu$m$^2$) & min. & 10.18 & min. & 14.84 & min. & 31.16 & min. & 298.00 \\
$t_f$ (sec)                       & \multicolumn{2}{c||}{13.81}  & \multicolumn{2}{c||}{20.68}  & \multicolumn{2}{c||}{18.07}  & \multicolumn{2}{c}{17.22}  \\

\hline


\rowcolor{gray!10}
\multicolumn{9}{c}{\textbf{OTA-4 (Low-voltage cascode OTA)}}\\
\hline
A$_{\mathrm{DC}}$ (dB)         & $\ge$20 & 21.37 & $\ge$30 & 39.57 & $\ge$40 & 41.58 & $\ge$60 & 60.03 \\
UGF (MHz)                       & $\ge$450 & 517.60 & max. & 15.33 & $\ge$60 & 68.71 & max. & 36.45 \\
BW$_{3\mathrm{dB}}$ (MHz)      & $\ge$30 & 48.19 & max. & 0.18 & $\ge$1 & 1.07 & max. & 0.08 \\
PM ($^\circ$)                   & $\ge$60 & 81.27 & $\ge$60 & 79.72 & $\ge$80 & 85.45 & $\ge$60 & 80.01 \\
CMRR (dB)                       & $\ge$50 & 54.53 & $\ge$50 & 58.51 & $\ge$50 & 61.62 & $\ge$55 & 60.52 \\
PSRR$+$ (dB)                 & max. & 29.81 & max. & 27.48 & max. & 31.58 & $\ge$30 & 35.46 \\
PSRR$-$ (dB)                 & max.& 31.01 & max. & 51.03 & max. & 53.04 & $\ge$40 & 43.08 \\
ICMR$_{\min}$ (V)               & $\le$0.4 & 0.34 & $\le$0.3 & 0.273 & $\le$0.4 & 0.27 & $\le$0.4 & 0.33 \\
ICMR$_{\max}$ (V)               & $\ge$0.7 & 0.71 & $\ge$0.7 & 0.77 & $\ge$0.7 & 0.78 & $\ge$0.7 & 0.80 \\
OCMR$_{\min}$ (V)               & $\le$0.2 & 0.18 & $\le$0.2 & 0.16 & $\le$0.2 & 0.16 & $\le$0.2 & 0.17 \\
OCMR$_{\max}$ (V)               & $\ge$0.7 & 0.82 & $\ge$0.7 & 0.79 & $\ge$0.7 & 0.81 & $\ge$0.7 & 0.82 \\
SR$+$ (V/$\mu$s)             & $\ge$300 & 303.67 & max. & 2.89 & $\ge$10 & 12.59 & max. & 32.58 \\
SR$-$ (V/$\mu$s)             & $\ge$300 & 359.71 & max. & 6.40 & $\ge$10 & 33.02 & max. & 87.72 \\
P$_{\mathrm{DC}}$ ($\mu$W)      & min. & 319.83 & $\le$8 & 7.87 & min. & 28.65 & min. & 112.40 \\
A$_{\mathrm{dev}}$ ($\mu$m$^2$) & min. & 20.71 & min. & 37.40 & min. & 183.60 & min. & 184.70 \\
$t_f$ (s)                        & \multicolumn{2}{c||}{22.60} & \multicolumn{2}{c||}{27.44}  & \multicolumn{2}{c||}{31.40}  & \multicolumn{2}{c}{17.49}  \\

\hline


\rowcolor{gray!10}
\multicolumn{9}{c}{\textbf{OTA-5 (Telescopic OTA)}}\\
\hline
A$_{\mathrm{DC}}$ (dB)         & $\ge$30 & 31.23 & $\ge$30 & 32.51 & $\ge$40 & 42.02 & $\ge$55 & 58.92 \\
UGF (MHz)                       & $\ge$400 & 417.22 & max. & 12.67 & $\ge$50 & 55.71 & max. & 15.33 \\
BW$_{3\mathrm{dB}}$ (MHz)       & $\ge$10 & 14.07 & max. & 0.08 & $\ge$1 & 1.23 & max. & 0.02 \\
PM ($^\circ$)                   & $\ge$70 & 78.10 & $\ge$60 & 81.23 & $\ge$80 & 82.11 & $\ge$80 & 85.53 \\
CMRR (dB)                       & $\ge$50 & 54.07 & $\ge$50 & 52.80 & $\ge$50 & 53.99 & $\ge$55 & 58.23 \\
PSRR$+$ (dB)                    & max. & 15.22 & max. & 17.36 & max. & 17.82 & $\ge$40 & 48.34 \\
PSRR$-$ (dB)                    & max. & 16.13 & max. & 22.35 & max. & 36.48 & $\ge$40 & 42.99 \\
ICMR$_{\min}$ (V)               & $\le$0.4 & 0.38 & $\le$0.3 & 0.25 & $\le$0.4 & 0.36 & $\le$0.4 & 0.38 \\
ICMR$_{\max}$ (V)               & $\ge$0.7 & 0.72 & $\ge$0.7 & 0.71 & $\ge$0.7 & 0.73 & $\ge$0.7 & 0.71 \\
OCMR$_{\min}$ (V)               & $\le$0.2 & 0.18 & $\le$0.2 & 0.18 & $\le$0.2 & 0.17 & $\le$0.2 & 0.16 \\
OCMR$_{\max}$ (V)               & $\ge$0.7 & 0.71 & $\ge$0.7 & 0.73 & $\ge$0.7 & 0.71 & $\ge$0.7 & 0.72 \\
SR$+$ (V/$\mu$s)             & $\ge$450 & 452.67 & max. & 5.18 & $\ge$10 & 17.05 & max. & 44.18 \\
SR$-$ (V/$\mu$s)             & $\ge$450 & 478.71 & max. & 6.74 & $\ge$10 & 21.24 & max. & 50.91 \\
P$_{\mathrm{DC}}$ ($\mu$W)      & min. & 938.44 & $\le$10 & 8.77 & min. & 94.25 & min. & 231.22 \\
A$_{\mathrm{dev}}$ ($\mu$m$^2$) & min. & 22.45 & min. & 85.96 & min. & 114.20 & min. & 310.95 \\
$t_f$ (s)                        & \multicolumn{2}{c||}{24.76} & \multicolumn{2}{c||}{33.23}  & \multicolumn{2}{c||}{28.21}  & \multicolumn{2}{c}{32.66}  \\

\hline
\end{tabular}%
}
\vspace{-3em}
\end{table}

\begin{table}[t]
\centering
\caption{GA-based COMP sizing.}
\label{tab:ga_comp_sizing}
\scriptsize
\setlength{\tabcolsep}{9pt}
\renewcommand{\arraystretch}{0.9}
\arrayrulecolor{black}

\resizebox{\columnwidth}{!}{%
\begin{tabular}{l|c|c||c|c}
\hline

\rowcolor{gray!18}
&
\multicolumn{2}{c||}{\textbf{StrongARM}}&
\multicolumn{2}{c}{\textbf{Double-Tail}}
\\

\hhline{~----}

\rowcolor{gray!18}
\multirow{-2}{*}{\textbf{Metric}} &
\textbf{Target} &
\textbf{SPICE} &
\textbf{Target} &
\textbf{SPICE}
\\
\hline

$t_{\text{eval}}$ (ps)                         & $\le$50 & 34.81 & $\le$50 & 44.36 \\
$t_{\text{pre}}$ (ps)                   & $\le$100 & 76.45 & $\le$40 & 38.57 \\
$E_{\text{dec}}$ (fJ)                        & min. & 193.29 & min. & 52.73 \\
ICMR$_{\min}$ (V)                  & $\le$0.4 & 0.34 & $\le$0.4 & 0.36 \\
ICMR$_{\max}$ (V)                  & $\ge$0.7 & 0.75 & $\ge$0.7 & 0.79 \\
A$_{\mathrm{dev}}$ ($\mu$m$^2$)   & min. & 4.66 & min. & 25.35 \\
$t_f$ (sec)                        & \multicolumn{2}{c||}{21.76} & \multicolumn{2}{c}{22.57} \\

\hline
\end{tabular}%
}
\vspace{-3mm}
\end{table}

\subsubsection{\textbf{Application-specific sizing}}

We employ GA to size the seven testcases in Table~\ref{tab:circuit_benchmarks_metrics_single} under application-specific performance objectives. For each OTA, we consider four sizing configurations: high speed, low power, balanced, and high gain. All four configurations use the same verified equations. The COMPs are sized under timing and input common-mode requirements, with energy and area minimization as additional objectives. Tables~\ref{tab:ga_ota_sizing} and~\ref{tab:ga_comp_sizing} present the target specifications alongside the final SPICE-verified results. Numerical specification bounds are enforced as hard constraints, while metrics marked ``min.'' or ``max.'' are treated as soft optimization objectives, following Section~\ref{sec:population_dse}. We additionally enforce transistor turn-on and saturation constraints for NMOS and PMOS devices:
$
|V_{\mathrm{GS}}|>|V_{\mathrm{TH}}|,
|V_{\mathrm{DS}}|>|V_{\mathrm{GS}}|-|V_{\mathrm{TH}}|
$.
These device-level constraints are omitted from the tables for brevity.

The OTA results in Table~\ref{tab:ga_ota_sizing} illustrate the tradeoffs associated with the four sizing configurations. High-speed sizing achieves higher UGF, BW$_{3\mathrm{dB}}$, and SR$+/-$ at increased power consumption, whereas low-power sizing reduces $P_{\mathrm{DC}}$ with lower speed. Balanced sizing addresses simultaneous gain, stability, signal-range, and speed requirements, while high-gain sizing increases $A_{\mathrm{DC}}$, generally with reduced bandwidth and increased $A_{\mathrm{dev}}$. For example, OTA-3 achieves a UGF of 583\,MHz at 1164\,$\mu$W in the high-speed configuration, compared with 28.4\,MHz at 18.6\,$\mu$W in the low-power configuration. Its high-gain configuration reaches 62.7\,dB, with an active-device area of 298\,$\mu$m$^2$.  For the comparators, Table~\ref{tab:ga_comp_sizing} shows that both designs satisfy their $t_{\mathrm{eval}}$, $t_{\mathrm{pre}}$, and ICMR specifications, with different outcomes for $E_{\mathrm{dec}}$ and $A_{\mathrm{dev}}$. Across all 22 application-specific sizing tasks, the selected designs satisfy every specification in the first SPICE-based verification. No additional sizing passes or specification tightening are required. Each optimization completes within 34\,s, followed by 7\,s for final verification.

\begin{table}[t]
\centering
\caption{Sizing efficiency across 100 specification sets.}
\label{tab:stage2_spec_stress}
\scriptsize
\setlength{\tabcolsep}{8pt}
\renewcommand{\arraystretch}{0.9}
\arrayrulecolor{black}

\resizebox{\columnwidth}{!}{%
\begin{tabular}{l|ccccc||cc}
\hline
\rowcolor{gray!18}
\textbf{Circuit} &
\multicolumn{5}{c||}{\textbf{Sizing attempts}} &
\multicolumn{2}{c}{\textbf{Average cumulative effort}} \\
\rowcolor{gray!18}
&
\textbf{1} & \textbf{2} & \textbf{3} &
\textbf{4} & \textbf{5} &
$\boldsymbol{\overline{t}_f}$ \textbf{(s)} &
$\boldsymbol{\overline{N}_e}$ \\
\hline

OTA-1  & 84 & 12 & 3 & 1 & 0 & 27.35 & 13553 \\
OTA-2  & 81 & 11 & 6 & 1 & 1 & 35.59 & 13991 \\
OTA-3  & 85 & 9 & 3 & 1 & 2 & 29.53 & 12882 \\
OTA-4  & 87 & 8 & 2 & 2 & 1 & 37.48 & 12177 \\
OTA-5  & 78 & 9 & 7 & 4 & 2 & 51.07 & 15254 \\
COMP-1 & 86 & 7 & 5 & 1 & 1 & 34.42 & 10541 \\
COMP-2 & 82 & 6 & 9 & 2 & 1 & 38.28 & 11991  \\

\hline
\end{tabular}%
}
\vspace{-5mm}
\end{table}

\begin{table}[t]
\centering
\caption{Representative spec-tightening recovery for OTA-3.}
\label{tab:stage2_tightening_example}

\scriptsize
\setlength{\tabcolsep}{0.8pt}
\renewcommand{\arraystretch}{1}
\arrayrulecolor{black}

\resizebox{\columnwidth}{!}{%
\begin{tabular}{
l|
c|c||
c|c||
c|c||
c|c}
\hline

\rowcolor{gray!18}
&
\multicolumn{2}{c||}{\textbf{Attempt 1}} &
\multicolumn{2}{c||}{\textbf{Attempt 2}} &
\multicolumn{2}{c||}{\textbf{Attempt 3}} &
&
\\

\hhline{
>{\arrayrulecolor{gray!18}}-
>{\arrayrulecolor{black}}|
--||--||--||
>{\arrayrulecolor{gray!18}}--
}

\rowcolor{gray!18}
\multirow{-2}{*}{\textbf{Metric}} &
\textbf{Target} & \textbf{SPICE} &
\textbf{Target} & \textbf{SPICE} &
\textbf{Target} & \textbf{SPICE} &
\multirow{-2}{*}{\shortstack{\textbf{Runtime }\\\textbf{$t_f$ (sec)}}} &
\multirow{-2}{*}{\textbf{$N_e$}}
\\
\arrayrulecolor{black}
\hline

A$_{\mathrm{DC}}$ (dB)      & \textcolor{green!50!black}{$\ge$25.0} & 34.11 & $\ge$25.0 & 30.49 & $\ge$25.0 & 29.71 & \multirow{13}{*}{\begin{tabular}{@{}c@{}}
\textbf{Attempt 1}: 19.8\\
\textbf{SPICE:} 6.9\\
+\\
\textbf{Attempt 2}: 41.3\\
\textbf{SPICE:} 6.9\\
+\\
\textbf{Attempt 3}: 43.9\\
\textbf{SPICE:} 6.9\\
=\\
\textbf{Total}\\
125.7
\end{tabular}} &
\multirow{13}{*}{\begin{tabular}{@{}c@{}}
\textbf{Attempt 1}\\
10201\\
+\\
\textbf{Attempt 2}\\
9600\\
+\\
\textbf{Attempt 3}\\
10280\\
=\\
\textbf{Total}\\
30081
\end{tabular}} \\

UGF (GHz)                   & \textcolor{green!50!black}{$\ge$14} & \textcolor{red}{12.6} & \textbf{$\ge$14.2} & 14.2 & $\ge$14.2 & 14.5 & & \\
BW$_{3\mathrm{dB}}$ (MHz)   & \textcolor{green!50!black}{$\ge$600} & \textcolor{red}{219.7} & \textbf{$\ge$700} & \textcolor{red}{383.6} & \textbf{$\ge$720} & 637.6 & & \\
PM ($^\circ$)               & \textcolor{green!50!black}{$\ge$60.0} & 89.57 & $\ge$60.0 & 89.53 & $\ge$60.0 & 89.50 & & \\
CMRR (dB)                   & \textcolor{green!50!black}{$\ge$50} & 51.34 & $\ge$50.0 & 57.88 & $\ge$50.0 & 52.06 & & \\
ICMR$_{\min}$ (V)           & \textcolor{green!50!black}{$\le$0.40} & \textcolor{red}{0.50} & \textbf{$\le$0.35} & \textcolor{red}{0.45} & \textbf{$\le$0.30} & 0.39 & & \\
ICMR$_{\max}$ (V)           & \textcolor{green!50!black}{$\ge$0.70} & 0.71 & $\ge$0.70 & 0.72 & $\ge$0.70 & 0.75 & & \\
OCMR$_{\min}$ (V)           & \textcolor{green!50!black}{$\le$0.10} & 0.03 & $\le$0.10 & 0.03 & $\le$0.10 & 0.03 & & \\
OCMR$_{\max}$ (V)           & \textcolor{green!50!black}{$\ge$0.60} & 0.62 & $\ge$0.60 & 0.66 & $\ge$0.60 & 0.66 & & \\
SR$+$ (V/$\mu$s)         & \textcolor{green!50!black}{$\ge$900} & \textcolor{red}{870.8} & \textbf{$\ge$950} & 990.9 & $\ge$950 & 989.7 & & \\
SR$-$ (V/$\mu$s)         & \textcolor{green!50!black}{$\ge$900} & \textcolor{red}{785.3} & \textbf{$\ge$950} & \textcolor{red}{847.0} & \textbf{$\ge$1000} & 904.5 & & \\
P$_{\mathrm{DC}}$ ($\mu$W)  & min. & 2450 & min. & 2060 & min. & 2491 & & \\
Area ($\mu\mathrm{m}^2$)    & min. & 36.98 & min. & 40.94 & min. & 48.50 & & \\

\hline
\end{tabular}%
}
\vspace{-3mm}
\end{table}

\subsubsection{\textbf{Specification-stress and recovery evaluation}}
\label{sec:stress_recovery}

\noindent
To evaluate specification satisfaction beyond the application-specific sizing tasks, we consider 100 distinct specification sets per testcase, totaling 700 sizing tasks. Table~\ref{tab:stage2_spec_stress} reports the number of attempts required to satisfy the specifications under SPICE-based verification. Of the 700 tasks, 583 (83.3\%) achieve feasibility on the first attempt, while the remaining tasks succeed within five attempts. The last two columns report the average cumulative runtime ($\overline{t}_f$) and candidate evaluations ($\overline{N}_e$) across all attempts until feasibility is achieved. Runtime includes optimization and verification for each attempt.

For tasks failing initial verification, we tighten the model-side targets of the failed metrics and repeat sizing without modifying the equations. Table~\ref{tab:stage2_tightening_example} illustrates this process for OTA-3. The green entries in Attempt~1 identify the original specifications, red values mark specification violations, and bold targets indicate subsequent tightening. The first attempt fails UGF, BW$_{3\mathrm{dB}}$, ICMR$_{\min}$, and SR$+/-$. After tightening, Attempt~2 satisfies UGF and SR$+$, while the other three metrics remain infeasible. Further tightening produces a design satisfying all original specifications in Attempt~3, requiring a cumulative runtime of 125.7\,s and 30,081 candidate evaluations. Feasibility is assessed against the original specifications, not the tightened model-side targets. For example, the final BW$_{3\mathrm{dB}}$ of 637.6\,MHz satisfies the original 600\,MHz requirement despite falling below the tightened 720\,MHz target. Thus, target tightening enables recovery from initial verification failures without regenerating the performance equations.

\begin{table}[t]
\centering
\caption{Optimizer portability with the equation-based performance model for sizing OTA-3.}
\label{tab:optimizer_comparison_2sota_passives}
\vspace{-1.5mm}
\scriptsize
\setlength{\tabcolsep}{3pt}
\renewcommand{\arraystretch}{0.82}

\resizebox{\columnwidth}{!}{%
\begin{tabular}{@{}lcccccccc@{}}
\hline
\rowcolor{gray!18}
\textbf{Opt.} &
\shortstack{\textbf{$A_{\mathrm{DC}}$}\\(dB)} &
\shortstack{\textbf{BW}\\(kHz)} &
\shortstack{\textbf{UGF}\\(MHz)} &
\shortstack{\textbf{CMRR}\\(dB)} &
\shortstack{\textbf{PSRR$^+$/PSRR$^-$}\\(dB)} &
\shortstack{\textbf{$P_{\mathrm{DC}}$}\\($\mu$W)} &
\shortstack{\textbf{Area}\\($\mu$m$^2$)} &
\shortstack{\textbf{$t_f$}\\(sec)} \\
\hline
GA  & 60.06 & 545.90 & 43.27 & 75.90 & 49.96/68.08 & 63.46 & 281.80 & 30.94 \\
PSO & 60.05 & 198.70 & 47.14 & 85.77 & 46.23/68.19 & 67.34 & 276.15 & 24.41 \\
DE  & 60.10 & 165.03  & 39.52  & 47.40 & 41.71/66.22 & 43.90 & 174.58 & 16.50 \\
\hline
\multicolumn{9}{c}{
$N_e=13{,}475$ per optimizer
($385$ candidates $\times$ $35$ generations)
} \\
\hline
\end{tabular}%

}

\vspace{-5mm}
\end{table}

\begin{table}[t]
\centering
\caption{Runtime speedup against SPICE-in-the-loop sizing.}
\label{tab:runtime_analysis}
\scriptsize
\setlength{\tabcolsep}{0.8pt}
\renewcommand{\arraystretch}{0.90}

\resizebox{\columnwidth}{!}{%
\begin{tabular}{@{}l l cc cc cc@{}}
\hline
\rowcolor{gray!18}
& &
\multicolumn{2}{c}{\textbf{GA}} &
\multicolumn{2}{c}{\textbf{PSO}} &
\multicolumn{2}{c}{\textbf{DE}} \\
\cline{3-8}

\rowcolor{gray!18}
\multirow[c]{-2}{*}{\textbf{Experiment}} &
\multirow[c]{-2}{*}{\textbf{Evaluation effort}} &
\textbf{Runtime} & \textbf{Speedup} &
\textbf{Runtime} & \textbf{Speedup} &
\textbf{Runtime} & \textbf{Speedup} \\
\hline

\shortstack[l]{SPICE-in-\\the-loop sizing}
&
\shortstack[l]{13,475 candidates}
&
112 min & n/a &
115 min & n/a &
114 min & n/a\\
\hline

\shortstack[l]{\textsc{NEMESIS2.0}-based \\sizing (One attempt)}
&
\shortstack[l]{
13,475 candidates \\
+ 1 SPICE verification
}
&
34.24 sec & 196.26$\times$ &
30.69 sec & 224.82$\times$ &
28.74 sec & 237.99$\times$
\\
\hline

\shortstack[l]{\textsc{NEMESIS2.0}-based \\sizing (Five attempts)}
&
\shortstack[l]{
67,375 candidates \\
+ 5 SPICE verifications
}
&
173.65 sec & 38.69$\times$ &
143.78 sec & 47.98$\times$ &
137.11 sec & 49.88$\times$
\\
\hline
\end{tabular}%
}

\vspace{-3mm}
\end{table}

\subsubsection{\textbf{Optimizer portability}}

\noindent
To evaluate optimizer portability, we apply GA, PSO, and DE to all seven testcases using the same verified performance equations. The optimizers share identical design variables, bounds, specifications, operating-region constraints, initial populations, and evaluation budgets. Each executes 35 generations with 385 candidates per generation, totaling $N_e=13{,}475$ evaluations. Table~\ref{tab:optimizer_comparison_2sota_passives} presents the resulting designs for a representative high-gain OTA-3 sizing task. GA achieves the highest BW$_{3\mathrm{dB}}$ (545.9\,kHz), PSO achieves the highest UGF (47.14\,MHz), while DE yields the lowest $P_{\mathrm{DC}}$ (43.90\,$\mu$W) and $A_{\mathrm{dev}}$ (174.58\,$\mu$m$^2$). These results show that the three optimizers produce different sizing outcomes using the same equation-based evaluator, without equation regeneration or optimizer-specific modifications.

\subsubsection{\textbf{Runtime analysis}}

\noindent
We compare \textsc{NEMESIS2.0}-based sizing with SPICE-in-the-loop sizing across all seven testcases. For each specification set and optimizer, both approaches use identical design variables, initialization, optimization settings, evaluation budgets, and computing resources. All runs complete the prescribed $13{,}475$ candidate evaluations, regardless of when feasibility is reached. SPICE-in-the-loop sizing executes metric-specific simulations for every candidate, with candidate evaluations parallelized across multiple CPU cores, whereas model-driven sizing evaluates the vectorized equations using extracted device parameters, followed by SPICE verification of the selected design. Table~\ref{tab:runtime_analysis} reports the mean runtimes and speedups. For a single sizing attempt, the SPICE-in-the-loop baseline requires 112--115\,min, compared with 28.74--34.24\,s for \textsc{NEMESIS2.0}, yielding speedups of 196.26--237.99$\times$.

The table also includes a five-attempt recovery scenario, which achieves speedups of 38.69--49.88$\times$ relative to the single-attempt SPICE-in-the-loop baseline. Model-generation runtime is excluded from these comparisons because the equations are generated once per circuit and reused across subsequent sizing tasks. Each equation-repair iteration requires 3--5\,min, primarily due to LLM latency and server load.
\vspace{-3mm}

\begin{table}[t]
\centering
\caption{Ablation study on circuit structure identification accuracy (\%) using netlist-only vs. netlist+schematic input.}
\label{tab:multimodal_ablation}

\scriptsize
\setlength{\tabcolsep}{1.8pt}
\renewcommand{\arraystretch}{1.03}
\arrayrulecolor{black}

\resizebox{\columnwidth}{!}{%
\begin{tabular}{l|cccc||cccc}
\hline
\rowcolor{gray!18}
&
\multicolumn{4}{c||}{\textbf{Netlist only}} &
\multicolumn{4}{c}{\textbf{Netlist + schematic}}
\\

\rowcolor{gray!18}
\multirow[c]{-2}{*}{\textbf{Circuit}} &
\textbf{Topology} &
\textbf{Primitives} &
\textbf{Roles} &
\textbf{Nets} &
\textbf{Topology} &
\textbf{Primitives} &
\textbf{Roles} &
\textbf{Nets}
\\
\hline

OTA-1  & 30.0 & 80.0 & 76.3 & 93.8 & 100.0 & 100.0 & 100.0 & 100.0 \\
OTA-2  & 50.0 & 86.7 & 86.0 & 86.7 & 100.0 & 100.0 & 100.0 & 100.0 \\
OTA-3  & 60.0 & 81.4 & 70.0 & 92.2 & 100.0 & 100.0 & 100.0 & 100.0 \\
OTA-4  & 70.0 & 82.0 & 78.3 & 88.9 & 100.0 & 100.0 & 100.0 & 100.0 \\
OTA-5  & 20.0 & 41.7 & 68.1 & 47.5 & 100.0 & 100.0 & 100.0 & 100.0 \\
COMP-1 & 80.0 & 87.1 & 77.0 & 86.3 & 100.0 & 100.0 & 100.0 & 100.0 \\
COMP-2 & 60.0 & 77.1 & 70.0 & 81.0 & 100.0 & 100.0 & 100.0 & 100.0 \\

\hline
\end{tabular}%
}
\vspace{-5mm}
\end{table}

\subsection{Importance of multimodality}
\label{sec:multimodal_input}

\noindent
We evaluate how schematic information affects the identification of circuit structures and device roles by the \textit{Topology Analysis Agent}. For each of the seven testcases, we perform 10 independent evaluations per input configuration, using either the netlist alone or the netlist with the schematic, while keeping all other inference settings identical. Table~\ref{tab:multimodal_ablation} compares the identification accuracy against ground-truth annotations across four categories: topology, primitives, device roles, and nets. Topology accuracy is the percentage of evaluations correctly identifying the circuit topology, while the remaining scores represent the average percentage of correctly identified primitives, device roles, and nets. The net-identification score includes circuit ports, signal paths, and connectivity.

With netlist-only input, topology identification accuracy ranges from 20\% to 80\%, with additional errors in primitive recognition, device-role assignment, and net interpretation. For example, OTA-3 achieves 60\% topology accuracy, 81.4\% primitive accuracy, 70.0\% role accuracy, and 92.2\% net accuracy. Incorporating the schematic increases all four scores to 100\% for OTA-3, with the same result observed across the remaining six testcases. Thus, although the netlist contains the required connectivity information, its textual representation alone does not consistently yield the correct circuit interpretation. The schematic provides additional structural context that eliminates the observed identification errors across the evaluated runs, establishing the benefit of multimodal inputs in circuit interpretation for performance equation generation.
\vspace{-4mm}

\section{Conclusion}
\label{sec:conclusion}

\noindent
This work presented \textsc{NEMESIS2.0}, an open-source agentic framework that automatically generates interpretable performance equations for analog circuits and applies them to automated sizing. The framework combines multimodal circuit interpretation, graph-based knowledge retrieval, and SPICE-guided equation repair to construct topology-specific analytical models without manual derivation. These equations are integrated with population-based optimization algorithms to size circuits under different performance objectives. Evaluated across five OTA and two COMP testcases in a commercial 65\,nm CMOS technology, the generated equations achieve errors within 5\% of SPICE. All 22 application-specific sizing tasks satisfy hard specifications on the first SPICE verification. Across 700 specification-stress tasks, 83.3\% achieve feasibility in the first attempt, with the remainder succeeding within five attempts through target tightening. For single-attempt sizing, the framework achieves 196--238$\times$ speedup over SPICE-in-the-loop sizing under identical evaluation budgets. These results show that automatically generated analytical equations enable efficient, application-specific analog sizing across circuit topologies and optimization algorithms while retaining independent SPICE verification.
\vspace{-3mm}

\section*{Acknowledgment}
\noindent
The framework incorporates LLMs, specifically OpenAI GPT-5.5, as a core modeling component, as detailed in Section~\ref{sec:NEMESIS2.0_modeling}. Beyond their role in the methodology, the authors used OpenAI ChatGPT to assist with the initial development of the software environment and with editorial revisions to improve the clarity and readability of the manuscript. All AI-assisted contributions were reviewed and verified by the authors.

\bibliographystyle{misc/IEEEtran}
\bibliography{bib/main.bib}


\makeatletter
\def\@IEEEBIOskipN{1.5\baselineskip}
\makeatother

\begin{IEEEbiography}[
{\includegraphics[width=1in,height=1.25in,
clip,keepaspectratio]{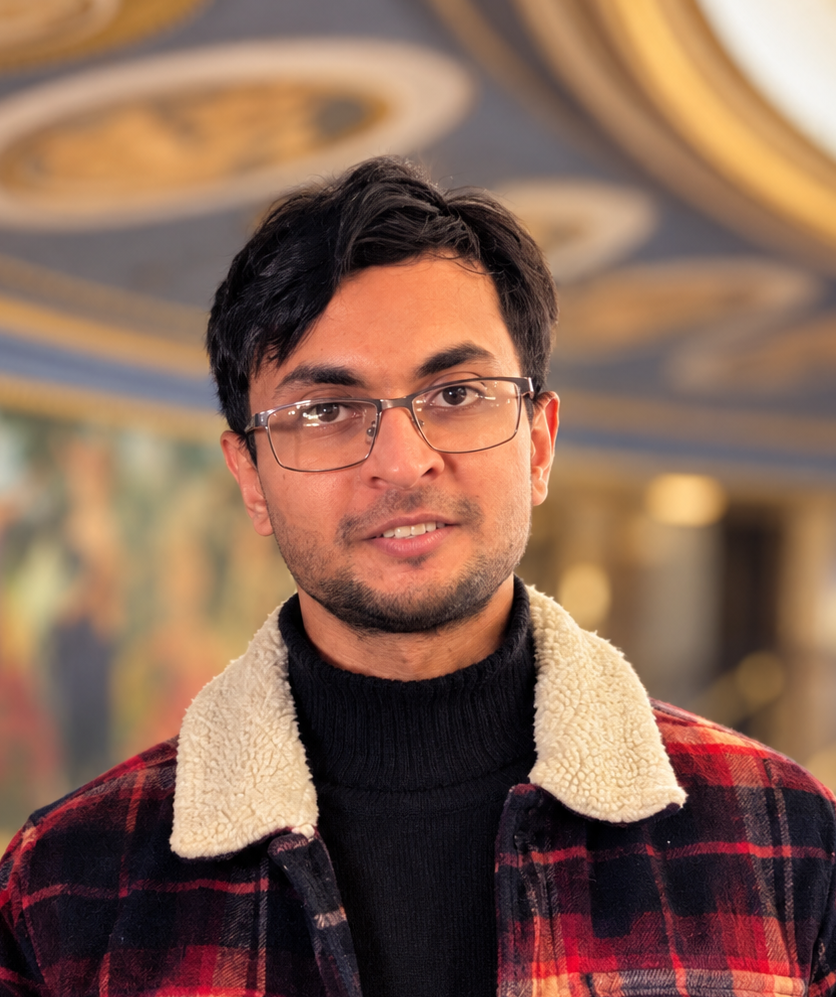}}
]{Subhadip Ghosh}
received the B.Tech. degree in electronics and telecommunication engineering from the Kalinga Institute of Industrial Technology, Bhubaneswar, India, in 2021, and the M.S. degree in electrical engineering from the University of Minnesota Twin Cities, Minneapolis, MN, USA, in 2025. He also pursued M.S. research at the Indian Institute of Technology Madras, Chennai, India, for one year. He is currently pursuing the Ph.D. degree in electrical engineering at the University of Minnesota Twin Cities. He was a Software Engineer Intern with Cadence Design Systems, San Jose, CA, USA, during the summers of 2023 and 2024. His research interests include analog/RF/mm-wave circuit design automation, circuit sizing and layout synthesis, and machine learning and generative AI techniques for electronic design automation.
\end{IEEEbiography}

\begin{IEEEbiography}[
{\includegraphics[width=1in,height=1.25in,
clip,keepaspectratio]{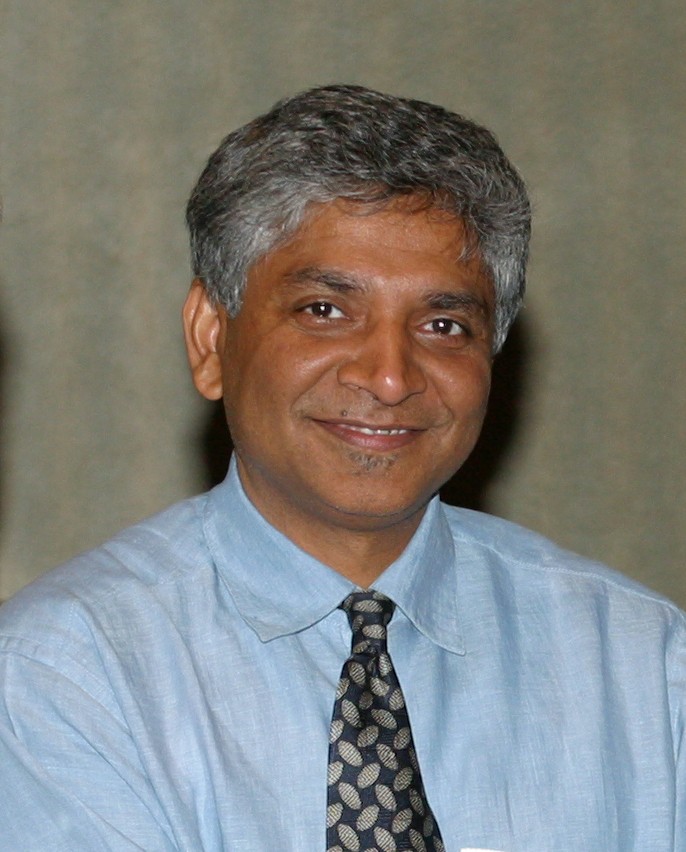}}
]{Ramesh Harjani}
received the B.S. degree from the Birla Institute of Technology and Science, Pilani, India, in 1982, the M.S. degree from the Indian Institute of Technology Delhi, India, in 1984, and the Ph.D. degree from Carnegie Mellon University, Pittsburgh, PA, USA, in 1989, all in electrical engineering. He is currently the Edgar F. Johnson Professor in the Department of Electrical and Computer Engineering at the University of Minnesota. Previously, he worked at Mentor Graphics Corporation, San Jose, CA, USA, and co-founded Bermai, Inc., a wireless multimedia CMOS chip startup, in 2001. He has also been a Visiting Professor at Lucent Bell Labs and the Army Research Laboratory. His research interests include analog/RF circuits for wired and wireless communications. He is a Fellow of the IEEE.
\end{IEEEbiography}

\begin{IEEEbiography}[
{\includegraphics[width=1in,height=1.25in,
clip,keepaspectratio]{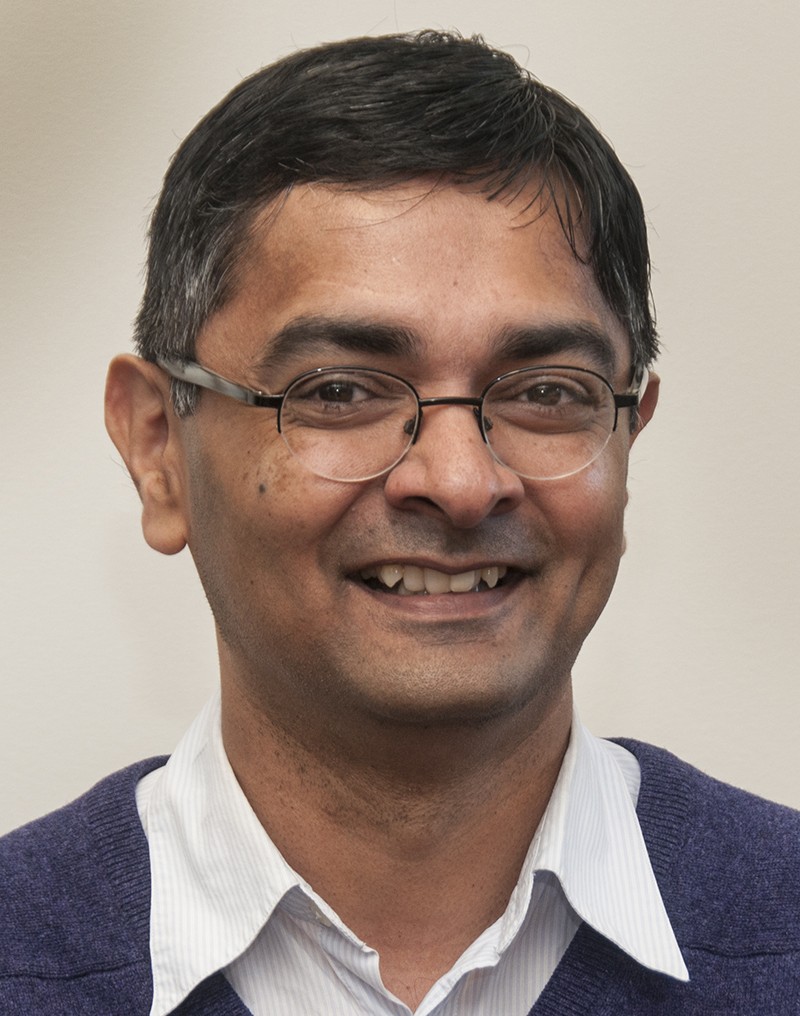}}
]{Sachin S. Sapatnekar}
received the B.Tech. degree from the Indian Institute of Technology Bombay, India, the M.S. degree from Syracuse University, NY, USA, and the Ph.D. degree from the University of Illinois at Urbana-Champaign, IL, USA. He is currently a Distinguished McKnight University Professor and holds the Robert and Marjorie Henle Chair at the University of Minnesota. His research interests include analog and digital design automation, machine learning for EDA, and emerging computing technologies. He has served as General Chair and Technical Program Chair of DAC, ISPD, and the Tau Timing Workshop, and as Editor-in-Chief of IEEE TCAD from 2010 to 2013. His honors include 12 Best Paper Awards, two ICCAD 10-Year Retrospective Most Influential Paper Awards, the SRC Technical Excellence Award (2003), the SIA University Research Award (2013), and the ACM SIGDA Pioneering Achievement Award (2026). He is a Fellow of the ACM and IEEE.
\end{IEEEbiography}

\end{document}